\documentclass[aps,prx,reprint,superscriptaddress,footnote, onecolumn,notitlepage,longbibliography]{revtex4-1}

\usepackage{relsize}
\usepackage{graphicx}
\usepackage{amsmath, bbm}
\usepackage{amssymb}
\usepackage{amsthm}
\usepackage{mathrsfs}
\usepackage{xcolor}
\usepackage{cancel}
\usepackage{titlesec}
\usepackage{enumitem}  
\usepackage[draft]{changes}

\usepackage[normalem]{ulem}

\titlespacing{\section}{0ex}{2ex}{0.4ex}
\titleformat{name=\section}{\bf}{\thesection.}{.3em}{}

\usepackage{soul}
\usepackage{dsfont}

\def\be{\begin{eqnarray}}
\def\ee{\end{eqnarray}}
\newcommand{\tr}[1]{\text{Tr}\left(#1\right)}
\newcommand{\Tr}[1]{\text{tr}\left(#1\right)}

\renewcommand{\>}{\rangle}

\theoremstyle{plain}

\newcommand{\blam}{\boldsymbol{\lambda}}

\begin{document}

\title{Thermodynamic length and work optimisation for Gaussian quantum states} 

\author{Mohammad Mehboudi}
\affiliation{D\' epartement de Physique Appliqu\' ee, Universit\' e de Gen\`eve, Gen\`eve, Switzerland}

\author{Harry J.~D. Miller}
\affiliation{Department of Physics and Astronomy, The University of Manchester, Manchester M13 9PL, UK.}

\date{\today}

\begin{abstract}
    Constructing optimal thermodynamic processes in quantum systems relies on managing the balance between the average excess work and its stochastic fluctuations. Recently it has been shown that two different quantum generalisations of thermodynamic length can be utilised to determine protocols with either minimal excess work or minimal work variance. These lengths measure the distance between points on a manifold of control parameters, and optimal protocols are achieved by following the relevant geodesic paths given some fixed boundary conditions. Here we explore this problem in the context of Gaussian quantum states that are weakly coupled to an environment and derive general expressions for these two forms of thermodynamic length. We then use this to compute optimal thermodynamic protocols for various examples of externally driven Gaussian systems with multiple control parameters.
\end{abstract}

\maketitle

\section{Introduction}
\

Microscopic systems that are driven out of equilibrium often exhibit noticeable fluctuations in their work output. Investigating the impact of these fluctuations has been a central topic in stochastic thermodynamics \cite{Seifert2012}, and this has led to a more detailed understanding of non-equilibrium thermodynamics via the celebrated fluctuation theorems \cite{Jarzynski1997c,Crooks,Campisi2011}. More recently, developments in quantum thermodynamics \cite{Vinjanampathy2015b} have led to a growing interest towards understanding how quantum mechanical effects impart signatures on the statistics of work \cite{Allahverdyan2014c,Baumer2018,Strasberg2019}, alongside the usual influence of classical stochastic fluctuations. A range of interrelated phenomena have been shown to affect the behaviour of work in the quantum regime, including quantum correlations \cite{Perarnau-Llobet2015b}, violations of macro-realism \cite{Solinas2018,Miller2018a}, contextuality \cite{Lostaglio2018}, quantum coherence \cite{Korzekwa2016,Brandner,Varizi2021,Francica2020} and quantum measurement effects \cite{Mohammady2019}.

One notable quantum signature that can be observed in the work statistics of an out-of-equilibrium process is the break down on the work fluctuation-dissipation relation (FDR) in the slow driving regime. For classical systems driven close to equilibrium, stochastic work is typically described by a normal distribution \cite{Speck}. In this case the average excess work above the free energy change, $\langle W_{\text{ex}} \rangle$, is proportional to the corresponding work variance $\langle \Delta W^2\rangle$ according to the FDR $\langle W_{\text{ex}} \rangle=\frac{1}{2}\beta\langle \Delta W^2\rangle$, with $\beta$ the inverse temperature of the surrounding environment \cite{Jarzynski1997c,Mandal2016a}. However, it has recently been proven that in an analogous quantum mechanical process that is driven slowly in time, this FDR can be violated and a more general relationship between the average and variance holds \cite{Miller2019,Scandi2019}:
\begin{align}\label{eq:qFDR}
    \langle W_{\text{ex}} \rangle=\frac{1}{2}\beta\langle \Delta W^2\rangle-\mathcal{I}_W,
\end{align}
Here $\mathcal{I}_W\geq 0$ is a measure of the non-commutativity between the system and the various conjugate forces acting upon it during the driving process. This measure is closely related to the Wigner-Yanase-Dyson skew information \cite{Wigner1963a}, which is commonly used as a measure of quantum fluctuations and uncertainty \cite{Luo2006}. The additional dissipation caused by these quantum fluctuations in the conjugate forces creates a non-classical signature in the overall work distribution, as this will typically deviate from a normal distribution \cite{Scandi2019}.

The breakdown of the work FDR has an immediate consequence if one is concerned with optimal thermodynamic control. Since the excess work is not generally proportional to the fluctuations, driving protocols with minimal excess work on average may not coincide with protocols with minimal work fluctuations. Managing this trade-off between fluctuations and dissipation is important for accurate and cost-effective control of quantum systems, and the equality~\eqref{eq:qFDR} demonstrates that non-commuting driving protocols may lead to increased irreversibility and reductions in reliability. However, if one wishes to transform a system from one equilibrium state to another in finite time, then  we cannot avoid the generation of quantum fluctuations if these two states do not commute. It is therefore necessary to take these limitations and trade-offs into account when trying to optimise a thermodynamic process in the quantum regime.

When restricting to regimes where a system is kept sufficiently close to equilibrium, the problem of finding paths with minimal excess work can be addressed using tools from thermodynamic geometry \cite{Ruppeiner1979,Salamon1983,Nulton2013,Crooks2007,Sivak2012a,Zulkowski2012}. It is possible to relate the average excess work to a Riemann metric tensor, or equivalently a form of thermodynamic length, which is subsequently minimised by following the associated geodesic within the manifold of control parameters \cite{Abiuso2020a}. This metric is closely related to the Fisher-Rao metric encountered throughout information geometry \cite{Burbea,Amari2007}. While classically this geodesic path will also minimise the work fluctuations, in the quantum case we know this will not generally hold due to the breakdown of the FDR. Instead, it was shown in \cite{Miller2019,Miller2020} that another notion of thermodynamic length must be minimised in order to reduce fluctuations. This length is constructed from an alternative metric that only becomes equivalent to the usual thermodynamic length in the limit of negligible quantum uncertainty in the conjugate forces. At present, not much is known about how these two different notions of thermodynamic length compare to each other in specific systems.

The goal of this paper is to determine these metrics for a wide class of quantum Gaussian systems and to use this to compute optimal work processes. Gaussian bosonic systems are widely employed in quantum information theory \cite{Weedbrook,Adesso2014a,Mehboudi} and quantum thermodynamics \cite{Brown2016c,Singh2019,Belenchia2020a} as they are experimentally accessible in current opto-mechanical experiments \cite{Zanin2019}. It is therefore valuable to understand how to optimise these kind of systems for thermodynamic tasks such as work extraction. We derive general analytic expressions for the two forms of thermodynamic length in open systems driven by a Gaussian Hamiltonian, whereby these quantities can be computed from knowledge of the system's covariance matrix and mean shift in the positions and momenta of the bosonic system. These general expressions provide a tool to find the protocols with minimal excess work or minimal fluctuations, and we illustrate this for three examples. The first example illustrates how our results connect to the classical limit, and we derive optimal protocols for a general open and classical Gaussian system with a single relaxation timescale. Our second example is a mean-shifted multi-mode Gaussian state in the fully quantum regime with a fixed covariance matrix. We show that in these two situations, protocols minimising the average excess work and variance are  equivalent. The third example we consider is a damped harmonic oscillator with a driven frequency and shift in mean position. In this case the two optimal protocols for the excess work and work variance are no longer equivalent, and we compare and contrast the geodesic paths associated with the two different metrics while demonstrating the relative merits of applying these protocols over naive ones.

\section{Excess work and fluctuations under slow driving: general case}

\

In this section we give a brief overview of how to describe the work statistics for a driven open quantum systems, along with how the resulting average work and variance can be connected to a pair of metric tensors in the slow driving regime. Let us first consider an open quantum system in contact with an environment whose local Hamiltonian $H(\blam_t)$ is parameterised by a set of $d$ scalar control variables, denoted by vector $\blam_t\in\mathbb{R}^d$ with
\begin{align}\label{eq:lam}
    \blam_t=\big\{\lambda_1(t),\lambda_2(t),...,\lambda_d(t)\big\}
\end{align}
We assume that our system is driven over a fixed duration $t\in[0,\tau]$ according to some path $\gamma:t\mapsto\blam_t$ in the parameter space given some fixed boundary conditions, namely
\begin{align}\label{eq:bound}
    \blam_0=\blam_A, \ \ \ \ \blam_\tau=\blam_B
\end{align}
Our second assumption is that the dynamics of this process can be modelled by a time-dependent Lindblad equation of generic form \cite{Albash2012}
\begin{align}\label{eq:lind}
    \dot{\rho}_t=\mathscr{L}_{\blam_t}[\rho_t]=-i[H(\blam_t),\rho_t]+\sum_{k=1}^D L_n(\blam_t) \rho_t L^\dagger_n(\blam_t)-\frac{1}{2}\{L_n^\dagger(\blam_t)L_n(\blam_t),\rho_t\},
\end{align}
where each of the $D$ jump operators $L_n(\blam_t)$ may depend on the control variables. We further assume that there exists a unique thermal fixed point at any given $\blam$, such that
\begin{align}\label{eq:thermal}
    \mathscr{L}_{\blam}[\pi(\blam)]=0; \ \ \ \ \ \pi(\blam)=\frac{e^{-\beta H(\blam)}}{\Tr{e^{-\beta H(\blam)}}},
\end{align}
with $\beta=1/(k_B T)$ the inverse temperature of the environment, and assume a thermal initial condition, $\rho_0=\pi(\blam_A)$. The equilibrium free energy of a thermal state is given by $F(\blam)=-\beta^{-1} \text{ln} \ \Tr{e^{-\beta H(\blam)}}$. For each control parameter we will define a corresponding conjugate observable that is shifted from its equilibrium average,
\begin{align}\label{eq:forces}
    X_j(\blam)=\frac{\partial H(\blam)}{\partial \lambda_j}-\tr{\pi_{\blam}\frac{\partial H(\blam)}{\partial \lambda_j}}. 
\end{align}
We will be interested in the excess work $W_{\text{ex}}$ along a given process, which is the difference between the work done and change in equilibrium free energy, $\Delta F=F_B-F_A$:
\begin{align}
     W_{\text{ex}} =W -\Delta F.
\end{align}
The work $W$ done on the system during a particular protocol is a stochastic variable that may be determined from the \textit{Two-Point-Measurement} protocol \cite{Talkner2007c}, where the system and environment are projected onto their respective energy eigenstates at the beginning and end of the process. This measurement scheme gives rise to a work distribution $P(W)$ that may be approximated in an appropriate weak coupling limit that is consistent with the approximations used to construct the Lindblad equation in~\eqref{eq:lind} \cite{Esposito2009,Silaev2014b}. Alternatively, for open systems this distribution may be accessed using quantum jump unravelling of the master equation, leading to an equivalent work distribution \cite{Horowitz2012,Horowitz2013b,Manzano2015,Liu2016b,Miller2021}. In our notation the average dissipation $\langle W_{\text{ex}}\rangle $ and work variance per unit temperature, $\langle \Delta W^2\rangle=\langle  W^2 \rangle-\langle  W \rangle^2$ are given by \cite{Suomela2014}
\begin{align}
    \label{eq:diss}\langle W_{\text{ex}} \rangle&=\int^\tau_0 dt \  \dot{\lambda}^j(t) \ \tr{X_j (\blam_t)\rho_t}, \\
    \label{eq:fluc}\langle \Delta W^2\rangle&=2\ \Re e\int^\tau_0 dt  \int^t_0 dt' \ \dot{\lambda}^k(t) \dot{\lambda}^j(t')\tr{X_k (\blam_t)\mathcal{P}(t,t')\big[\rho_{t'}\Delta_{\rho_{t'}}X_j (\blam_{t'})\big]}
\end{align}
where we adopt Einstein summation notation, denote the shifted observable $\Delta_{\rho_{t}}X_j (\blam_{t})=X_j (\blam_{t})-\tr{\rho_{t}X_j (\blam_{t})}$ and propagator $\mathcal{P}(t,t')=\overleftarrow{\text{exp}}\big(\int^t_{t'}d\nu \mathscr{L}_{\blam_\nu}\big)$. We will focus on the regime of slow driving, where the system stays close to its instantaneous thermal state at all times \cite{Cavina2017}. If the characteristic timescale of the system $\tau^{\text{eq}}$ is short compared to the total duration $\tau$, it has been shown that the average dissipation~\eqref{eq:diss} is approximated to linear order in $\tau^{\text{eq}}/\tau$ by the following expression \cite{Scandi}:
\begin{align}\label{eq:diss_slow}
    \<W_{\text{ex}}\>\simeq\int^\tau_0 dt \ \xi_{jk}(\blam_t) \  \dot{\lambda}^j(t)\dot{\lambda}^k(t),
\end{align}
where 
\begin{align}\label{eq:metric1}
    \xi_{jk}(\blam)=\frac{1}{2}\big(\tilde{\xi}_{jk}(\blam)+\tilde{\xi}_{kj}(\blam)\big)
\end{align}
is a symmetric tensor with elements
\begin{align}\label{eq:metric1_el}
    \tilde{\xi}_{jk}(\blam)= \int^\infty_0 d\nu \int^\beta_0 ds \ \tr{\mathscr{M}_{\blam}(\nu)\big[X_k(\blam)\big]\mathscr{U}_{\blam}(is)\big[X_j(\blam)\big] \ \pi_{\blam}}
\end{align}
Here we have defined a pair of evolution maps in the Heisenberg picture:
\begin{align}\label{eqq_evol_HeisenbergL}
&\mathscr{M}_{\blam}(\nu)[(.)]=e^{\nu \mathscr{L}^\dagger_{\blam}}[(.)], \\
&\mathscr{U}_{\blam}(\nu)[(.)]=e^{i \nu H(\blam)}[(.)] e^{-i \nu H(\blam)}.\label{eqq_evol_HeisenbergU}
\end{align}
Similarly, for the fluctuations~\eqref{eq:fluc} one can show that in the slow driving limit one has \cite{Miller2019}
\begin{align}\label{eq:fluc_slow}
    \langle \Delta W^2\rangle\simeq\int^\tau_0 dt \ \Lambda_{jk}(\blam_t) \dot{\lambda}^j(t)\dot{\lambda}^k(t),
\end{align}
where
\begin{align}\label{eq:metric2}
    \Lambda_{jk}(\blam)=\frac{1}{2}\big(\tilde{\Lambda}_{jk}(\blam)+\tilde{\Lambda}_{kj}(\blam)\big)
\end{align}
is another symmetric tensor with elements
\begin{align}\label{eq:metric2_el}
    \tilde{\Lambda}_{jk}(\blam)=2 \  \Re e\int^\infty_0 d\nu \ \tr{\mathscr{M}_{\blam}(\nu)[X_k(\blam)] X_j(\blam) \ \pi_{\blam}}.
\end{align}
If the conjugate observables commute with each other, this is sufficient to ensure the two tensors are proportional, meaning
\begin{align}
    [X_j(\blam), X_k(\blam)]=0 \ \ \forall j,k\implies \Lambda_{jk}(\blam)= 2k_B T \ \xi_{jk}(\blam).
\end{align}
If this is satisfied at points along a trajectory $\gamma$, one obtains the fluctuation-dissipation relation (FDR) \cite{Speck}:
\begin{align}\label{eq:FDR}
    \langle W_{\text{ex}} \rangle=\frac{1}{2}\beta\langle \Delta W^2\rangle.
\end{align}
However, in general the tensor elements $\xi_{jk}$ and $\Lambda_{jk}$ differ if there are pairs of non-commuting conjugate observables such as $[X_j(\blam),X_k(\blam)]\neq 0$. This leads to a breakdown of the FDR and we obtain the equality~\eqref{eq:qFDR} with a positive quantum correction $\mathcal{I}_W$, which can be viewed as a manifestation of quantum friction \cite{Feldmann}. If we additionally assume that the dynamics~\eqref{eq:lind} satisfy detailed balance, one may derive a matrix inequality \cite{Miller2019,Miller2021}
\begin{align}\label{eq:ineq}
    \Lambda\geq 2k_B T \ \xi\geq 0.
\end{align}
where $\xi$ and $\Lambda$ denote matrices with elements~\eqref{eq:metric1} and~\eqref{eq:metric2} respectively. The first inequality implies that the rate of excess work is less than half the rate of work fluctuations per unit temperature, 
\begin{align}\label{eq:FDR}
    \frac{d}{dt}\langle W_{\text{ex}} \rangle\leq\frac{1}{2}\beta\frac{d}{dt}\langle \Delta W^2\rangle,
\end{align}
with excess fluctuations stemming from non-commutativity of the set $\{X_j(\blam)\}$ relative to the equilibrium state. Note that this is a stronger inequality than that implied by the overall positivity of the quantum term $\mathcal{I}_W$ in~\eqref{eq:qFDR}. The second inequality means that the tensors $\xi$ and $\Lambda$ give rise to a semi-Riemann metric structure on the manifold of control parameters, because they are positive semi-definite, symmetric and smooth with respect to $\blam$ \cite{Amari2007}. For most situations of interest we can strengthen this condition so that $\xi$ and $\Lambda$ are positive definite, which is assumed throughout the rest of the paper. In this case the form of~\eqref{eq:diss_slow} and~\eqref{eq:fluc_slow} are that of a pair of action integrals acting on a Riemann manifold of the set of control parameters \cite{Sivak2012a,Zulkowski2012}. This means dissipation or fluctuations can be minimised using techniques from differential geometry that we outline in the next section. It is worth remarking that there may be weaker conditions than quantum detailed balance that ensure that the inequality~\eqref{eq:ineq} and associated metric structure holds, though this has not yet been established. 

\section{Work optimisation via the geodesic equation}

\

As already pointed out, the metric structure underlying the variables $\langle W_{\text{ex}} \rangle$ and $\langle \Delta W^2\rangle$ allows one to use a geometric approach to optimise the work dissipation or fluctuations of a slowly driven open quantum system. Optimal protocols are found by following a geodesic path along the space of control variables. In a fully quantum setting, there will be distinct geodesic paths that minimise either the fluctuations or dissipation due to the difference between the metrics $\xi$ and $\Lambda$. First note that both $\<W_{\text{ex}}\>$ and $\langle \Delta W^2\rangle$, as defined in~\eqref{eq:diss_slow} and~\eqref{eq:fluc_slow} respectively, take on the form of an action integral of generic form:
\begin{align}\label{eq:action}
    \mathcal{S}_{\gamma}=\int^\tau_0 dt \ g_{jk}(\blam_t) \dot{\lambda}^j(t)\dot{\lambda}^k(t),
\end{align}
where $\gamma:t\mapsto\blam_t$ denotes the protocol and $g_{jk}(\blam_t)$ a Riemann metric tensor. A protocol minimising the action will then satisfy the geodesic equation \cite{Amari2007}
\begin{align}\label{eq:geodesic}
    \frac{d^2 \lambda^i}{dt^2}+\Gamma^i_{jk}\frac{d\lambda^j}{dt}\frac{d\lambda^k}{dt}=0,
\end{align}
subject to the fixed boundary conditions~\eqref{eq:bound}. Here we denote the Christoffel symbols 
\begin{align}
    \Gamma^j_{ik}=
   \frac{1}{2} g^{jl}(\partial_{\lambda_i} g_{kl}+\partial_{\lambda_k} g_{il}-\partial_{\lambda_l} g_{ik}),
\end{align}
and $g^{jl}$ the metric inverse. The solution $\blam_t^{*}$ gives a minimal action $\mathcal{S}^{*}_{\gamma}$ proportional to the squared geodesic length,
\begin{align}
    \mathcal{S}_{\gamma}\geq \mathcal{S}^{*}_{\gamma}=\frac{\mathcal{L}^2}{\tau},
\end{align}
where the length traversed between the initial and final point is defined by
\begin{align}
    \mathcal{L}:=\int_0^\tau dt \sqrt{g_{jk}(\blam_t) \dot{\lambda}^j(t)\dot{\lambda}^k(t)}\bigg|_{\blam_t=\blam_t^{*}}
\end{align}
In our case we can minimise the average excess work by following a geodesic with respect to the metric $\xi$ in~\eqref{eq:metric1} and fixed boundary conditions~\eqref{eq:bound}. The minimum is then given by the corresponding squared length per unit time \cite{Sivak2012a,Scandi},
\begin{align}\label{eq:length1}
    \langle W_{\text{ex}} \rangle\geq \langle W_{\text{ex}} \rangle^*:=\frac{\mathcal{L}_{\mathcal{A}}^2}{\tau},
\end{align}
where
\begin{align}\label{eq:lengthAv}
    \mathcal{L}_{\mathcal{A}}:=\int_0^\tau dt \sqrt{\xi_{jk}(\blam_t) \dot{\lambda}^j(t)\dot{\lambda}^k(t)}\bigg|_{\blam_t=\blam_t^{\cal{A}}},
\end{align}
denotes our first notion of thermodynamic length, labelled with subscript $\mathcal{A}$ to indicate this quantity relates to the average work. We also denote $\blam_t^{\cal{A}}$ as the solution to the geodesic equation~\eqref{eq:geodesic} with respect to the metric choice $g=\xi$. Similarly, minimising the work fluctuations gives another geometric bound that is saturated by following a geodesic with respect to the other metric $\Lambda$ in~\eqref{eq:metric2}, so that
\begin{align}\label{eq:length2}
    \langle \Delta W^2\rangle\geq \langle \Delta W^2\rangle^*:=\frac{\mathcal{L}_{\mathcal{V}}^2}{\tau}
\end{align}
with 
\begin{align}\label{eq:lengthVar}
    \mathcal{L}_{\mathcal{V}}:=\int^\tau_0 dt \sqrt{\Lambda_{jk}(\blam_t) \dot{\lambda}^j(t)\dot{\lambda}^k(t)}\bigg|_{\blam_t=\blam_t^{\cal{V}}},
\end{align}
our second notion of thermodynamic length, which we label with $\mathcal{V}$ since this describes the variance rather than the average and solution $\blam_t^{\cal{V}}$ is determined from the geodesic equation~\eqref{eq:geodesic} by choosing metric $g=\Lambda$. Finding these two optimal paths via~\eqref{eq:geodesic} amounts to solving a set of $d$ coupled second order differential equations for each metric, which is typically a formidable task. However, we will demonstrate some examples that involve Gaussian open quantum systems where analytic solutions can be found. Before we consider solving any optimisation problems, in the next section we will first derive a general expression for the metric tensors~\eqref{eq:metric1} and~\eqref{eq:metric2} for quantum Gaussian states and processes.

\section{Metric tensors for quantum Gaussian systems}

\

We now focus specifically on $N$-mode bosonic systems with quadrature vector $R=(q_1,...,q_N, p_1,...,p_N)^T$, with $q_n$ and $p_n$ the respective position and momentum operator for the $n$'th mode. The quadrature operators satisfy a bosonic algebra 
\begin{align}
    [R_n,R_m]=\Omega_{nm},
\end{align}
where $\Omega_{nm}$ the symplectic form ($\hbar=1$), which is a $2N\times 2N$ matrix with block form
\begin{align}
    \Omega:=i\left(\begin{array}{cc}
          \mathbb{O}_N & \mathbb{I}_N   \\ 
          -\mathbb{I}_N  & \mathbb{O}_N  
    \end{array}\right).
\end{align}
We consider a class of Hamiltonians that are at most quadratic with respect to the quadrature $R$, which can be expressed in the form
\begin{align}\label{eq:gauss}
    H_{\blam} = \frac{1}{2}R_{\blam}^T G_{\blam}R_{\blam}.
\end{align}
Here we have introduced a symmetric  matrix $G_{\blam}\in\mathbb{R}^{2N\times 2N}$ representing the second order coupling terms in the Hamiltonian, while $R_{\blam}=R-\mu_{\blam}$ a linear order shift in the quadratures with real vector $\mu_{\blam}\in\mathbb{R}^{2N}$. We allow for external control of both the quadratic and linear terms in the Hamiltonian, and thus the matrix $G_{\blam}$ and vector $\mu_{\blam}$ depend on a set of $d$ control parameters $\blam$ as defined by~\eqref{eq:lam}. The quadratic Hamiltonian is such that the corresponding thermal state $\pi_{\blam}$ in~\eqref{eq:thermal} is a quantum Gaussian state \cite{Braunstein2005,Weedbrook}. From this we can express the conjugate forces~\eqref{eq:forces} in the form
\begin{align}
    X_j(\blam)=\frac{1}{2}\Tr{\mathcal{X}_j(\blam) (\Sigma_{\blam}-\sigma_{\blam})} - \big(x_j(\blam)\big)^T G_{\blam} R_{\blam},
\end{align}
where we denote $[\Sigma_{\blam}]_{nm}=\frac{1}{2}\{[R_{\blam}]_n,[R_{\blam}]_m\}$ as the matrix of second order quadratures with respect to $R_{\blam}$, while $\mathcal{X}_j(\blam)=\partial_{\lambda_j}G_{\blam}$ and $x_j(\blam)=\partial_{\lambda_j} \mu_{\blam}$ are the respective derivatives of the quadratic and linear terms in the Hamiltonian with respect to the $j$th control variable. Note that in the above notation we use a lower case $'\text{tr}'$ to indicate a trace over the real matrix space $\mathbb{R}^{2N\times 2N}$ rather than the Hilbert space. Finally, we also denote the thermal covariance matrix by
\begin{align}\label{eq:covmat}
    \sigma_{\blam} = \tr{\Sigma_{\blam} \pi_{\blam}}=\frac{1}{2}\coth \big(\beta G_{\blam} \Omega/2\big)\Omega.
\end{align}
which contains all the second moments of the system with respect to the steady state $\pi_{\blam}$ \cite{Chen2005,Banchi2015}. We also note that for a thermal state $\pi_{\blam}$, the mean quadrature (also known as the displacement vector) equals the linear shift, i.e., $ \tr{\pi_{\blam} R}=\mu_{\blam}$. As we consider an open system, we further assume that the Lindblad jump operators in~\eqref{eq:lind} are linear in quadratures, namely
\begin{align}\label{eq:jump}
    L_n(\blam):=c^T_n(\blam)R_{\blam},
\end{align}
where $c_n(\blam)\in\mathbb{C}^{2N}$ is a vector of complex numbers that may depend on the control parameters. Equations \eqref{eq:gauss} and \eqref{eq:jump} guarantee that dissipative dynamics is Gaussian, that is it maps Gaussian state to Gaussian states \cite{Mehboudi}.

Next, let us focus on the evolution maps~ \eqref{eqq_evol_HeisenbergL} and~ \eqref{eqq_evol_HeisenbergU} in the Gaussian formalism. It suffices to characterise the evolution of the first and second order quadrature operators that read
\begin{align}
    {\mathscr M}_{\blam}(\nu)(\Sigma_{\blam}) &=F_{{\blam},\nu} \Sigma_{\blam} F_{{\blam},\nu}^T + \int_0^{\nu} d\nu^{\prime} F_{{\blam},\nu^{\prime}}   D_{\blam} F_{{\blam},\nu^{\prime}}^T,\label{eq:Gauss_diss_quad} \\
    {\mathscr M}_{\blam}(\nu)(R_{\blam}) &
    = F_{{\blam},\nu}R_{\blam},
\end{align}
where $F_{{\blam},\nu}=e^{\nu A_{\blam}}$, $A_{\blam}=-i\Omega\big(G_{\blam}-\text{Im}(C^\dagger_{\blam}C_{\blam})\big)$ and $D_{\blam}=\Omega \text{Re}(C^\dagger_{\blam}C_{\blam})\Omega $ \cite{Mehboudi}. Here we define the rectangular matrix $C_{\blam}=(c^T_1(\blam);...;c^T_D(\blam))^T\in\mathbb{C}^{2N\times D}$ . The application of the channel to $\sigma_{\blam}$ should be understood through its application on the identity operator, since in fact by $\sigma_{\blam}$ we mean $\sigma_{\blam} {I}$. Therefore, ${\mathscr M}_{\blam}(\nu)(\sigma_{\blam}) = \sigma_{\blam}$ due to unitality of the dynamics. Furthermore, by using the fact that $\sigma_{\blam}$ represents the covariance matrix of the fixed point of the dynamics, by multiplying both sides of \eqref{eq:Gauss_diss_quad} with $\pi_{\blam}$ and taking the trace, we have
\begin{align}
    \sigma_{\blam} =
    {\mathscr M}_{\blam}(\nu)(\sigma_{\blam}) =   F_{{\blam},\nu} \sigma_{\blam} ~ F_{{\blam},\nu}^T + \int_0^{\nu} d\nu^{\prime} F_{{\blam},\nu^{\prime}}   D_{\blam} F_{{\blam},\nu^{\prime}}^T \ \ \ \ \forall \nu\in[0,\infty),
\end{align}
which by putting together with \eqref{eq:Gauss_diss_quad} gives 
\begin{align}
    {\mathscr M}_{\blam}(\nu)(\Sigma_{\blam}-\sigma_{\blam}) = F_{{\blam},\nu} (\Sigma_{\blam} - \sigma_{\blam}) ~ F_{{\blam},\nu}^T.
\end{align}
As for the unitary map in~\eqref{eqq_evol_HeisenbergU}, we have
\begin{align}
    {\mathscr U}_{\blam}(\nu)(\Sigma_{\blam} - \sigma_{\blam}) &= e^{i \nu H_{\blam}}(\Sigma_{\blam} - \sigma_{\blam}) e^{-i \nu H_{\blam}} = S_{G_{\blam}}^{\nu}\Sigma_{\blam} S_{G_{\blam}}^{\nu T} - \sigma_{\blam},\\
    {\mathscr U}_{\blam}(\nu)(R_{\blam}) &= e^{i \nu H_{\blam}} R_{\blam} e^{-i \nu H_{\blam}} = S_{\blam}^{\nu}R_{\blam}
\end{align}
where we use the Baker–Campbell–Hausdorff lemma, and define $S_{\blam}^{\nu}= e^{-i\nu\Omega G_{\blam}}$. By substituting in~\eqref{eq:metric1_el}, one gets
\begin{align}\label{eq_integrand_xi}
    \nonumber\tilde{\xi}_{jk}(\blam) &=\int^\infty_0 d\nu \int^\beta_0 ds \  \tr{\mathscr{M}_{\blam}(\nu)[X_{j}(\blam)]\mathscr{U}_{\blam}(is)[X_{k}(\blam)] \ \pi_{\blam}}, \\ 
    &=\int^\infty_0 d\nu \int^\beta_0 ds \  \text{Tr}\bigg[\bigg(\frac{1}{2}\Tr{(F_{{\blam},\nu}(\Sigma_{\blam} - \sigma_{\blam}) F_{{\blam},\nu}^T) \mathcal{X}_j(\blam)} - x^T_j(\blam) G_{\blam} F_{\blam,\nu}R_{\blam}\bigg)\nonumber\\
    & \ \ \ \ \ \ \  \ \ \ \ \ \ \ \ \ \ \ \ \ \ \ \ \ \ \ \ \ \ \ \ \ \ \ \ \ \ \ \ \ \times \bigg(\frac{1}{2}\Tr {S_{\blam}^{is}(\Sigma_{\blam} - \sigma_{\blam}) ~(S_{\blam}^{is })^T\mathcal{X}_k(\blam)} - x_k^T(\blam)G_{\blam} S_{\blam}^{is}R_{\blam}\bigg)\pi_{\blam}\bigg],\nonumber\\
    & 
    =\nonumber\int^\infty_0 d\nu \int^\beta_0 ds \  \bigg(\frac{1}{2}\Tr{(S_{\blam}^{is })^T \mathcal{X}_k(\blam) S_{\blam}^{is} \big((\sigma_{\blam}-\frac{1}{2}\Omega) F_{{\blam},\nu}^T \mathcal{X}_j(\blam) F_{{\blam},\nu}(\sigma_{\blam}+\frac{1}{2}\Omega)\big)} \\
    & \ \ \ \ \ \ \  \ \ \ \ \ \ \ \ \ \ \ \ \ \ \ \ \ \ \ \ \ \ \ \ \ \ \ \ \ \ \ \ \  +x_j^T(\blam) G_{\blam}F_{\blam,\nu}(\sigma_{\blam} + \frac{1}{2}\Omega) S_{\blam}^{ix T}  G_{\blam}x_k(\blam) \bigg),
\end{align}
where we used \textit{Wick's theorem} in order to expand the fourth order correlations in terms of second moments. To simplify our notation we now define the following maps:
\begin{align}
    &\mathcal{J}_{\blam}[(.)]:=\int^\beta_0 ds \ (S_{\blam}^{is})^T (.)S_{\blam}^{is}, \\
    &\mathcal{F}_{\blam}[(.)]:=\int^\infty_0 d\nu \ F_{{\blam},\nu}^T (.)F_{{\blam},\nu},
\end{align}
and the matrix
\begin{align}
    Y_{\blam}=\int^\infty_0 d\nu \ F_{{\blam},\nu}.
\end{align}
We then have 
\begin{align}
    \nonumber\tilde{\xi}_{jk}(\blam)&=\frac{1}{2}\Tr{\mathcal{J}_{\blam}[ \mathcal{X}_k(\blam)]\big(\sigma_{\blam}-\frac{1}{2}\Omega\big) \mathcal{F}_{\blam} [\mathcal{X}_j(\blam)]\big(\sigma_{\blam}+\frac{1}{2}\Omega\big)}+ x_j^T(\blam) G_{\blam}Y_{\blam}(\sigma_{\blam} + \frac{1}{2}\Omega) \bigg(\int^\beta_0 ds  \big[e^{s\Omega G_{\blam}}\big]^T\bigg)  G_{\blam}x_k(\blam), \\
    \nonumber
    &=\frac{1}{2}\Tr{\mathcal{J}_{\blam}[ \mathcal{X}_k(\blam)]\big(\sigma_{\blam}-\frac{1}{2}\Omega\big) \mathcal{F}_{\blam} [\mathcal{X}_j(\blam)]\big(\sigma_{\blam}+\frac{1}{2}\Omega\big)}+ x_j^T(\blam) G_{\blam}Y_{\blam}(\sigma_{\blam}\Omega + \frac{\mathbb{I}}{2}) \bigg(\int^\beta_0 ds   \ \Omega\big[e^{s\Omega G_{\blam}}\big]^T\Omega\bigg)\Omega  G_{\blam}x_k(\blam), \\
    \nonumber
    &=\frac{1}{2}\Tr{\mathcal{J}_{\blam}[ \mathcal{X}_k(\blam)]\big(\sigma_{\blam}-\frac{1}{2}\Omega\big) \mathcal{F}_{\blam} [\mathcal{X}_j(\blam)]\big(\sigma_{\blam}+\frac{1}{2}\Omega\big)}+ x_j^T(\blam) G_{\blam}Y_{\blam}(\sigma_{\blam}\Omega + \frac{\mathbb{I}}{2}) \bigg(\int^\beta_0 ds  \big[e^{-s\Omega G_{\blam}}\big]\bigg)\Omega G_{\blam}x_k(\blam), \\
    \nonumber
    &=\frac{1}{2}\Tr{\mathcal{J}_{\blam}[ \mathcal{X}_k(\blam)]\big(\sigma_{\blam}-\frac{1}{2}\Omega\big) \mathcal{F}_{\blam} [\mathcal{X}_j(\blam)]\big(\sigma_{\blam}+\frac{1}{2}\Omega\big)}+ x_j^T(\blam) G_{\blam}Y_{\blam}(\sigma_{\blam}\Omega + \frac{\mathbb{I}}{2})\big(\mathbb{I}-e^{-\beta \Omega G_{\blam}}\big)\big(\Omega G_{\blam} \big)^{-1}\Omega G_{\blam}x_k(\blam) \\
    &=\frac{1}{2}\Tr{\mathcal{J}_{\blam}[ \mathcal{X}_k(\blam)]\big(\sigma_{\blam}-\frac{1}{2}\Omega\big) \mathcal{F}_{\blam} [\mathcal{X}_j(\blam)]\big(\sigma_{\blam}+\frac{1}{2}\Omega\big)}+ x_j^T(\blam) G_{\blam}Y_{\blam}(\sigma_{\blam}\Omega + \frac{\mathbb{I}}{2})\big(\mathbb{I}-e^{-\beta \Omega G_{\blam}}\big)x_k(\blam)
\end{align}
where in the second line we used $\Omega^2=\mathbb{I}$ with $\mathbb{I}$ being the $2N\times 2N$ dimensional identity matrix (not to be mistaken with $I$ which is the identity operator in the infinite dimensional Hilbert space), in the third line we used $\Omega f(A)\Omega=f(\Omega A \Omega)$ for any function $f(x)$ and matrix $A$, and in the fourth line we evaluated the integral with respect to $s$. The second term can be simplified by using the Cayley transform:
\begin{align}
    e^{-\beta\Omega G_{\blam}}=(2\sigma_{\blam}\Omega-\mathbb{I})(2\sigma_{\blam}\Omega+\mathbb{I})^{-1}.
\end{align}
which follows from the expression for covariance matrix in~\eqref{eq:covmat}. This gives us a final expression for the tensor
\begin{align}\label{eq:metric_gauss1}
    \tilde{\xi}_{jk}(\blam)=\frac{1}{2}\Tr{\mathcal{J}_{\blam}[ \mathcal{X}_k(\blam)]\big(\sigma_{\blam}-\frac{1}{2}\Omega\big) \mathcal{F}_{\blam} [\mathcal{X}_j(\blam)]\big(\sigma_{\blam}+\frac{1}{2}\Omega\big)}+ x_j^T(\blam) G_{\blam}Y_{\blam}x_k(\blam).
\end{align}
In a similar fashion we evaluate the other tensor~\eqref{eq:metric2}. This gives
\begin{align}
    \nonumber\tilde{\Lambda}_{jk}(\blam)&=2 \  \Re e\int^\infty_0 d\nu \ \tr{\mathscr{M}_{\blam}(\nu)[X_j(\blam)] X_i(\blam) \ \pi_{\blam}}, \\
    & =2{\Re e}\int_0^{\infty}d\nu {\rm Tr}\bigg[\bigg(\frac{1}{2}\Tr{(F_{{\blam},\nu}(\Sigma_{\blam} - \sigma_{\blam}) F_{{\blam},\nu}^T) \mathcal{X}_j(\blam)}- x^T_j(\blam)G_{\blam} F_{\blam,\nu}R_{\blam}\bigg) \nonumber\\
    & \ \ \ \ \ \ \ \ \ \ \ \  \ \ \  \ \ \ \ \ \ \ \  \ \ \ \ \ \ \ \ \ \ \ \ \ \ \ \ \ \ \ \ \ \ \ \ \ \ \ \  \times \bigg(\frac{1}{2}\Tr {(\Sigma_{\blam} - \sigma_{\blam}) \mathcal{X}_k(\blam)}- x_k^T(\blam) G_{\blam}R_{\blam}\bigg)\pi_{\blam}\bigg]\nonumber, \\
    & = {\Re e}~\int_0^{\infty}d\nu\bigg(\Tr{ \mathcal{X}_k(\blam)  \big((\sigma_{\blam}-\frac{1}{2}\Omega) F_{{\blam},\nu}^T \mathcal{X}_j(\blam) F_{{\blam},\nu}(\sigma_{\blam}+\frac{1}{2}\Omega)\big)} + 2 x_j^T(\blam)G_{\blam} F_{\blam,\nu}(\sigma_{\blam} + \frac{1}{2}\Omega) G_{\blam}x_k(\blam)\bigg).
\end{align}
This can be written in the final form
\begin{align}\label{eq:metric_gauss2}
    \tilde{\Lambda}_{jk}(\blam)={\Re e}~\Tr{ \mathcal{X}_k(\blam)\big(\sigma_{\blam}-\frac{1}{2}\Omega\big) \mathcal{F}_{\blam}[\mathcal{X}_j(\blam)]\big(\sigma_{\blam}+\frac{1}{2}\Omega\big)}+2 ~ x_j^T(\blam) G_{\blam}Y_{\blam}\sigma_{\blam}   G_{\blam}x_k(\blam).
\end{align}
The expressions for the tensors~\eqref{eq:metric_gauss1} and~\eqref{eq:metric_gauss2} now allow us to compute the thermodynamic metrics directly from the variables composing the set $(G_{\blam},\{c_k (\blam)\},\mu_{\blam})$ which appear in the Lindblad equation. Note that in both instances the metric decomposes into a sum between one contribution dependent on the second order coupling terms $\{\mathcal{X}_j(\blam)\}$, and another contribution from the linear terms $\{x_j(\blam)\}$. This means that if the quadratic and linear terms in the Hamiltonian are controlled independently, then the metric tensors $\xi(\blam)$ and $\Lambda(\blam)$ will split into a block diagonal form with linear and quadratic terms separated. As we will demonstrate with some examples, this simplifies the computation of geodesics through the manifold.

\newpage

\section{Computing geodesic paths}

\

In this final section of the paper we present three different examples of Gaussian open systems that can be optimised using our geometric methods.

\

\noindent \textit{Example 1: Open system in the classical limit.} 

\

\noindent It is instructive to consider what happens to our expressions in the classical limit. To construct this limit we need to reintroduce a factor of $\hbar$ that we had previously neglected for convenience, which follows by the replacement $\Omega\to\hbar\Omega$. We can define a classical limit by treating $\hbar\ll 1$ as small and Taylor expanding the covariance matrix~\eqref{eq:covmat} \cite{Tanaka2006}, which in the limit $\hbar\to 0$ gives
\begin{align}
\lim_{\hbar\to 0}\sigma_{\blam} = \lim_{\hbar\to 0} \ \frac{\hbar}{2}\coth \big(\beta \hbar G_{\blam} \Omega/2\big)\Omega=k_B T \ G^{-1}_{\blam}
\end{align}
Applying this expansion to the excess work metric~\eqref{eq:metric_gauss1} gives
\begin{align}\label{eq:metric_class1}
    \nonumber\lim_{\hbar\to 0}\tilde{\xi}_{jk}(\blam)&=\frac{1}{2}\lim_{\hbar\to 0}\int^\beta_0 ds \ \Tr{ \big[e^{is\hbar\Omega G_{\blam}}\big]^T \mathcal{X}_k(\blam) e^{-is\hbar\Omega G_{\blam}}\big(\sigma_{\blam}-\frac{\hbar}{2}\Omega\big) \mathcal{F}_{\blam} [\mathcal{X}_j(\blam)]\big(\sigma_{\blam}+\frac{\hbar}{2}\Omega\big)}+   x_j^T(\blam) G_{\blam}Y_{\blam}x_k(\blam), \\
    &=\frac{1}{2}k_B T \ \Tr{ \mathcal{X}_k(\blam) G^{-1}_{\blam}\mathcal{F}_{\blam} [\mathcal{X}_j(\blam)]G^{-1}_{\blam}}+  x_j^T(\blam) G_{\blam}Y_{\blam}x_k(\blam).
\end{align}
Similarly, taking the limit $\hbar\to 0$ for the work fluctuations~\eqref{eq:metric_gauss2} gives
\begin{align}\label{eq:metric_class2}
    \lim_{\hbar\to 0}\tilde{\Lambda}_{jk}(\blam)=(k_B T)^2 \ \Tr{ \mathcal{X}_k(\blam) G^{-1}_{\blam}\mathcal{F}_{\blam} [\mathcal{X}_j(\blam)]G^{-1}_{\blam}}+2 k_B T \ x_j^T(\blam) G_{\blam}Y_{\blam}x_k(\blam).
\end{align}
Comparing these two expressions tells us that the two metrics are proportional for $\hbar\ll 1$, since 
\begin{align}
    \Lambda\simeq 2k_B T \xi.
\end{align}
This means that we recover the work FDR~\eqref{eq:FDR} in this classical limit, so that
\begin{align}\label{eq:class_lim}
    \langle W_{\text{ex}} \rangle\simeq \frac{1}{2}\beta\langle \Delta W^2\rangle.
\end{align}
By treating $\hbar\ll 1$ we are essentially guaranteeing that all conjugate forces associated with the system Hamiltonian will commute with the thermal state $\pi_{\blam}$ at zeroth order in $\hbar$, which means quantum fluctuations can be neglected and the term $\mathcal{I}_W$ in~\eqref{eq:qFDR} becomes negligible. Our results are therefore consistent with classical derivations of the work FDR that model the open system with Focker-Planck dynamics, such as \cite{Speck}. In this classical regime the metric tensor $\xi$ is closely related the Fisher-Rao metric over the manifold of Gaussian probability distributions. To see this consider an equation of motion with a single characteristic timescale $\tau^{\text{eq}}$ that is independent of the control parameters \cite{Scandi,Abiuso2020a}, such as
\begin{align}
    \mathscr{L}_{\blam_t}[\rho_t]=\frac{\pi(\blam_t)-\rho_t}{\tau^{\text{eq}}}.
\end{align}
In this case one can show that the classical metric~\eqref{eq:metric_class1} reduces to
\begin{align}\label{eq:sigel}
    \xi_{jk}(\blam)=\frac{1}{2}\big(\tilde{\xi}_{jk}(\blam)+\tilde{\xi}_{kj}(\blam)\big)=\frac{1}{2  }\tau^{\text{eq}} k_B T \ \Tr{ \mathcal{X}_k(\blam) G^{-1}_{\blam}\mathcal{X}_j(\blam) G^{-1}_{\blam}}+ \tau^{\text{eq}} x_j^T(\blam) G_{\blam}x_k(\blam),
\end{align}
It then follows that $\beta\xi(\blam)/\tau^{\text{eq}}$ is equivalent to the Fisher-Rao metric for a Gaussian distribution with covariance matrix $k_B T G^{-1}_{\blam}$ and mean vector $\mu_{\blam}$ \cite{Burbea,Amari2007}. In this case analytic expressions for geodesic paths are well-known. For example, consider the case where $\mu_{\blam}=0$ and we can control all elements of the covariance matrix. Then the excess work per $k_B T$ and variance can be compactly expressed as
\begin{align}
    \beta \langle W_{\text{ex}} \rangle=\frac{1}{2}\beta^2\langle \Delta W^2\rangle=\frac{1}{2}\tau^{\text{eq}}\int^\tau_0 dt \ \Tr{\big(G_{\blam}^{-1}\frac{d G_{\blam}}{dt}\big)^2},
\end{align}
The integrand is proportional to the squared line element $ds^2$ of a Siegel metric \cite{Burbea,Lenglet2006}, defined by
\begin{align}\label{eq:siegel}
    ds:=\sqrt{\frac{1}{2}\Tr{\big(G_{\blam}^{-1} \ d G_{\blam}\big)^2}},
\end{align}
such that we can write
\begin{align}
    \langle W_{\text{ex}} \rangle=k_B T \ \tau^{\text{eq}} \int_0^\tau \bigg(\frac{ds}{dt}\bigg)^2,
\end{align}
Using~\eqref{eq:length1} the excess work can be lower bounded $\langle W_{\text{ex}} \rangle\geq\langle W_{\text{ex}} \rangle^*$ according to
\begin{align}
    \langle W_{\text{ex}} \rangle^*= k_B T\bigg(\frac{\tau^{\text{eq}}}{\tau}\bigg)\mathcal{L}^2
\end{align}
where 
\begin{align}
    \mathcal{L}=\sqrt{\frac{1}{2}\Tr{\log^2 G^{-1/2}_{\blam_A} G_{\blam_B} G^{-1/2}_{\blam_A} }},
\end{align}
is the geodesic length between an initial and final Gaussian thermal state \cite{Burbea}. This bound is tight and saturated by following the geodesic curve $\blam^*_t$ such that
\begin{align}
    G_{\blam^*_t}=G_{\blam_A}^{1/2}\exp \bigg(\frac{t}{\tau}  \log \big(G^{-1/2}_{\blam_A} G_{\blam_B} G^{-1/2}_{\blam_A}\big) \bigg) G_{\blam_A}^{1/2}.
\end{align}
A proof of this fact can be found in \cite{Andai2009a}. Note that clearly due to the validity of the work FDR the optimal protocol minimises both the average excess work and variance, ie. $\blam^*_t=\blam^{\mathcal{A}}_t=\blam^{\mathcal{V}}_t$. For more complicated open classical systems where there are multiple relaxation timescales associated with the environment, general expressions for the geodesics are not known since one has to consider the modifications to the Siegel line element~\eqref{eq:siegel} that come from the integral relaxation time of the environment \cite{Feldmann2013,Sivak2012a}. There are some examples of more complicated Gaussian classical systems that can be solved, such as a driven one-dimensional harmonic potential \cite{Zulkowski2012}.

\

\noindent \textit{Example 2: Driving via linear coupling.} 

\

\noindent Returning to a fully quantum example, consider a situation where we treat the displacement vector components, $\blam=\mu_{\blam}$, as our control variables while fixing the quadratic terms in the matrix $G$. We further assume the Lindblad jump operators are linear with respect to the quadratures, meaning
\begin{align}
H_{\blam} = \frac{1}{2}R_{\blam}^T G R_{\blam}, \ \ \ \ \ \ \ \ \ \ \ L_n(\blam):=c^T_n R_{\blam}.
\end{align}
In this situation, the equilibrium free energy will not change, namely $\Delta F=0$. This is because for a Gaussian Hamiltonian, the free energy is independent of the linear coupling $\mu_{\blam}$. As a result, processes of this form will always consume work and hence
\begin{align}
    \langle W \rangle \geq 0.
\end{align}
In terms of optimisation, we are interested in minimising the average work cost and fluctuations required to displace a thermal state with initial mean $\mu_{\blam_A}$ to a new thermal state with mean $\mu_{\blam_B}$ over a finite but slow process. From our formulae~\eqref{eq:metric_gauss1} and~\eqref{eq:metric_gauss2} we can express the metric tensors in the matrix form
\begin{align}
    &\xi=\frac{1}{2} G Y+Y^T G, \\
    &\Lambda= G \bigg(Y\sigma +\sigma Y^T\bigg)G,
\end{align}
where again we write $Y_{\blam}=Y$ as this does not depend on $\blam$. Since the above metrics are independent of the control parameters, we immediately see that the geodesic path satisfying~\eqref{eq:geodesic} will be a linear protocol with 
\begin{align}\label{eq:lin}
    \blam_t^*=\frac{(\tau-t)\blam_A+t\blam_B}{\tau}.
\end{align}
Crucially, this optimal solution will simultaneously minimise both the dissipation and fluctuations uniquely so that, as we found in the classical limit, $\blam^*_t=\blam^{\mathcal{A}}_t=\blam^{\mathcal{V}}_t$. This corresponds in both cases to a flat manifold with vanishing curvature, and the minimal work done work and minimal variance are then given by the squared lengths~\eqref{eq:lengthAv} and~\eqref{eq:lengthVar} with
\begin{align}
    \langle W \rangle^* &=\frac{1}{\tau}\xi_{jk}(\lambda_B^j-\lambda_A^j)(\lambda_B^k-\lambda_A^k), \\
    \langle \Delta W^2 \rangle^* &=\frac{1}{\tau}\Lambda_{jk}(\lambda_B^j-\lambda_A^j)(\lambda_B^k-\lambda_A^k).
\end{align}
The corresponding geodesic lengths can be expressed in terms of the initial and final displacement vectors, such that 
\begin{align}
    &\mathcal{L}_{\mathcal{A}}=\sqrt{(\mu_{\blam_B}-\mu_{\blam_A})^T \xi (\mu_{\blam_B}-\mu_{\blam_A}) }, \\
    &\mathcal{L}_{\mathcal{V}}=\sqrt{(\mu_{\blam_B}-\mu_{\blam_A})^T \Lambda (\mu_{\blam_B}-\mu_{\blam_A}) },
\end{align}
which may be interpreted as the Euclidean distances between the images of $\mu_{\blam_B}$ and $\mu_{\blam_A}$ under the respective transformations $\mu\mapsto P\mu$ and $\mu\mapsto \tilde{P}\mu$, where $\xi=P P^T$ and $\Lambda=\tilde{P} \tilde{P}^T$ are the Cholesky decompositions of the two metrics \cite{Distributions2020}. Our result here demonstrates that the cheapest and most reliable way to displace a Gaussian open quantum system, close to equilibrium, is via a naive protocol. Interestingly, this example demonstrates that there can be situations where the optimal protocols minimising $\langle W_{\text{ex}} \rangle$ and $\langle \Delta W^2 \rangle$ can still coincide even though the work FDR is broken due to non-commuting conjugate forces (ie. the quantum term $\mathcal{I}_W$ in~\eqref{eq:qFDR} is positive definite). 

\

\noindent \textit{Example 3: Driving a damped harmonic oscillator.} 

\

\noindent So far we have only encountered situations where the optimal protocol uniquely minimise the average excess work and fluctuations. However, in the quantum regime this is most often not the case, as we demonstrate in this final example. We consider a single harmonic oscillator of unit mass weakly coupled to a thermal bath, whose frequency $\omega$ and mean position $y$ can be controlled externally. The Gaussian Hamiltonian~\eqref{eq:gauss} is composed of the terms
\begin{align}
    G_{\omega}=\left(\begin{array}{cc}
          \omega^2 & 0   \\ 
          0 & 1  
    \end{array}\right), \ \ \ \ \  \mu^T_{y}=(y,0).
\end{align}
Note that we have used a change of notation $G_{\blam}\to G_{\omega}$ and $\mu_{\blam}\to \mu_y$ to indicate that the collective control variables are
\begin{align}
    \blam_t:=\big\{\omega(t),y(t)\big\}.
\end{align}
We assume there are two jump operators for the Lindblad equation~\eqref{eq:lind} which we label as
\begin{align}
    L_1(\blam)=c^T_1(\omega) R_{y}, \ \ \ \ \ \ \ L_2(\blam)=c^T_2(\omega) R_{y},
\end{align}
where
\begin{align}
    &c_{1}^{T}(\omega)=\frac{1}{2}\sqrt{\gamma(\omega)} \ \big(1, \ \ i/\omega\big), \\
    & c_{2}^{T}(\omega)=\frac{1}{2}\sqrt{\gamma(-\omega)} \ \big(1, \ \ -i/\omega\big),
\end{align}
Here we denote the damping rate by
\begin{align}
\gamma(\omega)= 2J(\omega)(N(\omega) + 1)
\end{align}
where $N(\omega) = [\exp(\omega/T) - 1]^{-1}$ is the average occupation of a bosonic mode with frequency $\omega$ at temperature $T$, the spectral density is $J(\omega)=~\gamma_0~\omega$ and $\gamma_0$ represents the coupling strength between the system and bath. The thermal covariance matrix is given by
\begin{align}\label{eq:fixed_single_osc}
\sigma_{\omega} = \frac{1}{2}\coth(\beta\omega/2)\left(
\begin{array}{cc}
1/\omega & 0\\
0 & \omega 
\end{array}
\right).
\end{align}
Introducing the matrix $C_{\omega}=(c^T_1(\omega);c^T_2(\omega))^T$, we calculate the following:
\begin{align}
\nonumber A_\omega &=-i\Omega\big(G_{\omega}-\text{Im}(C^\dagger_{\omega}C_{\omega})\big), \\
&=\left(
\begin{array}{cc}
-\gamma_0/2 & 1 \\
-\omega^2 & -\gamma_0/2
\end{array}
\right), 
\end{align}
Taking the exponential gives
\begin{align}
\nonumber F_{{\omega},\nu}&=\exp(\nu A_\omega) \\
&= e^{-\gamma_0\nu/2}\left(
\begin{array}{cc}
\cos(\omega\nu) & \sin(\omega\nu)/\omega \\
-\omega\sin(\omega\nu) & \cos(\omega\nu)
\end{array} \right)
\end{align}
Furthermore we have 
\begin{align}\label{eq:SG_single_osc}
\nonumber S_\omega^{ix} &=e^{x\Omega G_{\omega}}, \\
&=\left(
\begin{array}{cc}
\cosh(\omega x) & i\sinh(\omega x)/\omega \\
-i\omega\sinh(\omega x) & \cosh(\omega x)
\end{array}
\right).
\end{align}
To compute the metric $\xi$, we simply plug these matrices into the expression~\eqref{eq:metric_gauss1}, and after some lengthy but straightforward algebra we obtain a diagonal tensor
\begin{align}\label{eq:xi_1}
    \xi_{11} &= \frac{ \omega ~{\rm csch}^2(\beta\omega/2) + \frac{2\gamma_0^2\coth(\beta\omega/2)}{\gamma_0^2+4\omega^2}}{16\gamma_0\omega^3}\simeq \frac{1}{16 \omega^2}~{\rm csch}^2(\beta \omega/2)\bigg(\frac{\beta }{\gamma_0}+\frac{\gamma_0}{2\omega^3}\sinh(\beta \omega/2)\cosh(\beta \omega/2)\bigg), \\
        \label{eq:xi_2}\xi_{22}& = \frac{\gamma_0 \omega^2}{2\gamma_0^2 + 8\omega^2}\simeq \frac{\gamma_0}{8}, \\
        \xi_{12}&=0.
\end{align}
where we have taken an approximation $\gamma^2\ll \omega^2$ which is consistent with weak coupling used to derive the original Lindblad equation. A similar calculation for the other metric~\eqref{eq:metric_gauss2} yields
\begin{align}
    \label{eq:lam_1}\Lambda_{11} &\simeq\frac{1}{4 \omega^2}~{\rm csch}^2(\beta \omega/2)\bigg(\frac{ 1}{\gamma_0}+\frac{\gamma_0}{4\omega^2}\cosh(\beta \omega)\bigg), \\
        \label{eq:lam_2}\Lambda_{22}&=\simeq\frac{\gamma_0 \omega}{4}\coth(\beta\omega/2), \\
        \Lambda_{12}&=0.
\end{align}
As a consistency check, one can verify the inequality $\Lambda_{jj}\geq 2k_B T \ \xi_{jj}\geq 0$ for $j=1,2$ as expected by~\eqref{eq:ineq}. We also recover the equality $\Lambda\simeq 2k_B T \ \xi$ in the high temperature limit $\beta^2\omega^2\ll 1$, which implies the classical work FDR $\langle W_{\text{ex}}\rangle =\frac{1}{2}\beta \langle \Delta W^2 \rangle+\mathcal{O}(\beta^2\omega^2)$. We can view this as an alternative form of classical limit~\eqref{eq:class_lim}, where in this case the thermal fluctuations are much larger than any quantum contributions. For general temperatures, we now turn to computing the geodesic paths corresponding to minimal excess work and minimal work fluctuations. Unlike our previous example, in this case these geodesic paths will be distinct from each other due to the difference between the metric tensors. Concerning the metric $\xi$, we find that there is only one non-zero Christoffel symbol given by
\begin{align}
    \Gamma^1_{11}=\partial_\omega \log \sqrt{\xi_{11}[\omega]}, 
\end{align}
where we have switched notation $\xi_{11}\to \xi_{11}[\omega]$ to highlight the dependence of the metric on the frequency. Substituting this into the geodesic equation~\eqref{eq:geodesic} gives us a pair of decoupled second order differential equations:
\begin{align}\label{eq:geo_om}
    &\frac{d^2 \omega}{dt^2}+\bigg(\frac{d\omega}{dt}\bigg)^2\partial_\omega \log \sqrt{\xi_{11}[\omega]}=0, \\
    &\frac{d^2 y}{dt^2}=0, 
\end{align}
The solution $\blam^{ {\cal A}}_t = \{\omega^{\cal{A}}(t),y^{\cal{A}}(t)\}$ yields a rate of change for each parameter given by
\begin{align}\label{eq:geodesic_ex1}
    &\frac{d}{dt}\omega^{\cal{A}}(t)=\frac{\big(\omega_B-\omega_A\big)\xi_{11}^{-1/2}[\omega^{\cal{A}}(t)]}{\int^\tau_0 dt' \ \xi_{11}^{-1/2}[\omega^{\cal{A}}(t')]}, \\
    &\frac{d}{dt}y^{\cal{A}}(t)=\frac{y_B-y_A}{\tau} \label{eq:geodesic_ex2}
\end{align}
This means that in order to minimise the excess work, one needs to vary the average position at a constant rate while changing the frequency at a rate proportional to the inverse square root of the friction $\xi_{11}$.

\begin{figure*}[htbp!]
\begin{center}
\includegraphics[width=1\textwidth]{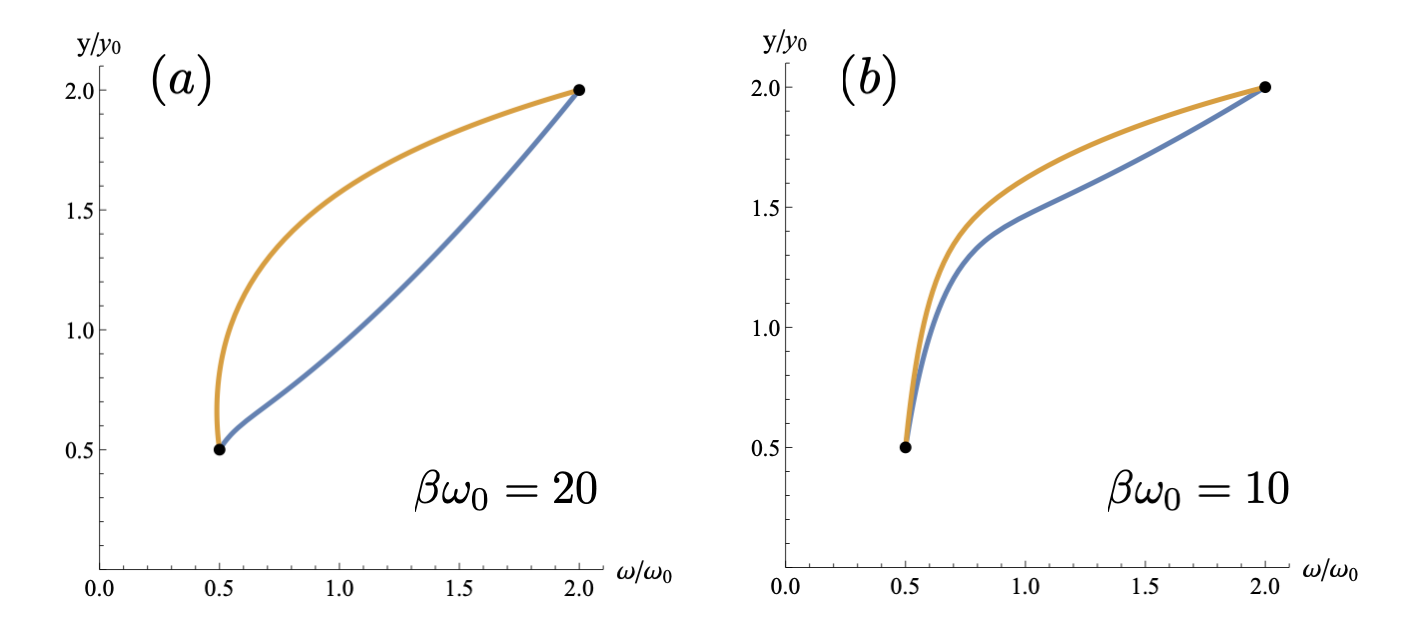}
\vspace*{-0.2cm}
\caption{Plot of different geodesic solutions for the damped oscillator with boundary conditions $\blam_A=\{0.5\omega_0,0.5y_0 \}$ and $\blam_B=\{2\omega_0, 2 y_0 \}$ with parameters $\beta\omega_0=20$, $\tau=100/\omega_0$ and $\gamma_0=0.1 \omega_0$, with $\omega_0=y_0=1$ a reference frequency and position. The blue curve represents the geodesic solution to~\eqref{eq:geodesic_ex1} and~\eqref{eq:geodesic_ex2} that minimises the average excess work, while the orange curve represents a geodesic for the alternative equations~\eqref{eq:geo_om1} and~\eqref{eq:geo_om2} that minimise the work fluctuations. The temperatures plotted are \textbf{(a)} $\beta\omega_0=20$ and \textbf{(b)} $\beta\omega_0=10$.}
\vspace*{-0.5cm}
\label{fig:geodesic}
\end{center}
\end{figure*}

To find the geodesic for the metric $\Lambda$ we cannot find a closed form expression since the resulting geodesic equation remains coupled between the parameters $\omega(t)$ and $y(t)$. To see this, first note that we now have four non-zero Christoffel symbols given by
\begin{align}
    &\Gamma^1_{11}=\partial_\omega \log \sqrt{\Lambda_{11}[\omega]}, \\
    &\Gamma^1_{22}=-\frac{1}{2}\Lambda_{11}^{-1}[\omega] \ \partial_\omega \Lambda_{22}[\omega], \\
    &\Gamma^2_{1 2}=\Gamma^2_{21}=\partial_\omega \log \sqrt{\Lambda_{22}[\omega]}.
\end{align}
Substituting this into the geodesic equation gives us a pair of coupled second order differential equations:
\begin{align}\label{eq:geo_om1}
    &\frac{d^2 \omega}{dt^2}+\bigg(\frac{d\omega}{dt}\bigg)^2\partial_\omega \log \sqrt{\Lambda_{11}[\omega]}-\frac{1}{2}\bigg(\frac{dy}{dt}\bigg)^2\Lambda_{11}^{-1}[\omega] \partial_\omega \Lambda_{22}[\omega]=0, \\
    &\frac{d^2 y}{dt^2}+2\bigg(\frac{d\omega}{dt}\bigg)\bigg(\frac{dy}{dt}\bigg)\partial_\omega \log \sqrt{\Lambda_{22}[\omega]}=0, \label{eq:geo_om2}
\end{align}
Since these equations are coupled, they cannot be solved independently of each other, unlike the solution for minimal excess work. One immediate consequence of this is that a naive protocol that changes the mean position at a constant rate will in fact not minimise the work fluctuations. In Figure 1 we plot the geodesic solutions optimising either the excess work (ie. solution ~\eqref{eq:geodesic_ex1} and~\eqref{eq:geodesic_ex2}) or the fluctuations (ie. numerical solution to ~\eqref{eq:geo_om1} and~\eqref{eq:geo_om2}) in the parameter space. It is clear that these paths are distinct from each other, with a larger discrepancy shown at lower temperatures as we expect due to the breakdown of the work fluctuation-dissipation relation~\eqref{eq:FDR}.

\begin{figure*}[htbp!]
\begin{center}
\includegraphics[width=1\textwidth]{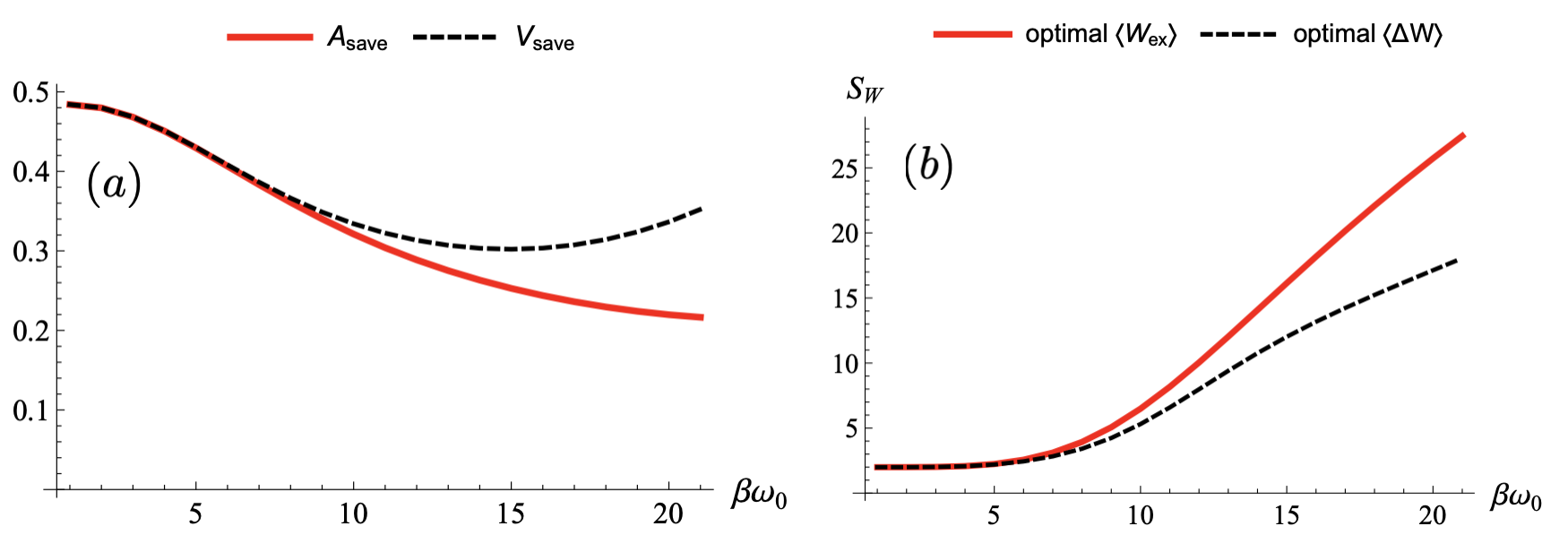}
\vspace*{-0.2cm}
\caption{Plot \textbf{(a)} shows the quantities $A_{\text{save}}$ defined in~\eqref{eq:Asave} and $V_{\text{save}}$ defined in~\eqref{eq:Vsave} as a function of inverse temperature. Plot \textbf{(b)} displays the Fano factor $S_W$ in~\eqref{eq:fano} for the two different geodesic paths, where the red line corresponds to the minimal excess work protocol while the black line corresponds to the minimal fluctuation protocol. In both figures the boundary conditions are $\blam_A=\{0.5\omega_0,0.5y_0 \}$ and $\blam_B=\{2\omega_0, 2 y_0 \}$ with $\gamma_0=0.1 \omega_0$ and $\tau=100$.}
\vspace*{-0.5cm}
\label{fig:geodesic}
\end{center}
\end{figure*}

In order to quantify the benefit of choosing a geodesic path as opposed to a naive protocol, we introduce the following quantities:
\begin{align}\label{eq:Asave}
    A_{\text{save}}:=\frac{\langle W_{\text{ex}} \rangle^{*}}{\langle W_{\text{ex}} \rangle_{\text{lin}}},
\end{align}
\begin{align}\label{eq:Vsave}
    V_{\text{save}}:=\frac{\langle \Delta W^2 \rangle^{*}}{\langle \Delta W^2  \rangle_{\text{lin}}}.
\end{align}
Here $\langle W_{\text{ex}} \rangle^{*}$ denotes the minimal excess work~\eqref{eq:length1} done while following the geodesic protocol~\eqref{eq:geodesic_ex1} and~\eqref{eq:geodesic_ex2}, while $\langle W_{\text{ex}} \rangle_{\text{lin}}$ is the same quantity when following a linear protocol $\blam_t=\big((\tau-t)\blam_A+t\blam_B\big)/\tau$. Similarly, $\langle \Delta W^2 \rangle^{*}$ is the minimal work variance~\eqref{eq:length2} determined by the geodesic path satisfying ~\eqref{eq:geo_om1} and~\eqref{eq:geo_om2}, and is compared to the amount $\langle \Delta W^2  \rangle_{\text{lin}}$ obtained by a linear protocol.  In Figure 2 \textbf{(a)} we plot these quantities as a function of the inverse temperature. For the parameter values shown in the figure, one finds improvements of up to $80\%$ for the excess work, while up to $70\%$ improvements in the work fluctuations. It is worth noting that neither $A_{\text{save}}$ or $V_{\text{save}}$ are monotonic with respect to temperature, meaning that one may obtain more significant improvements over naive protocols in both dissipation and fluctuations at intermediate temperatures. Nevertheless, clearly one can see that the geodesic paths have the potential to provide very significant improvements over naive protocols. Given that we observe distinct geodesic protocols for minimising the excess work versus the fluctuations, it is also worth considering how much fluctuations or dissipation can increase if one follows the opposite geodesic. To quantify this, we turn to the dimensionless Fano factor:
\begin{align}\label{eq:fano}
    S_W:= \frac{\langle \Delta W^2 \rangle}{k_B T \langle W_{\text{ex}} \rangle}.
\end{align}
This measures the dispersion of the work probability distribution, and we plot its value in Figure 2 \textbf{(b)} as a function of inverse temperature. It can be seen that in general, dispersion increases at lower temperatures, while approaches the expected value $S_W=2$ in the high temperature limit in accordance with the work FDR~\eqref{eq:FDR}. If one follows the geodesic with minimal excess work, we generally cause the work distribution to become more dispersed than if we follow the geodesic minimising the fluctuations, and this increase occurs with a greater slope for this choice of protocol as we approach lower temperatures.

\section{Conclusions}

\

We have considered the problem of finding optimal protocols that minimise both the excess work and work fluctuations for general Gaussian open quantum systems subject to external driving and fixed boundary conditions. By addressing this problem in the regime of slow driving, where the system remains close to an instantaneous thermal state at all times, we were able to apply geometric techniques to find these optimal protocols. We derived a set of general expressions for the corresponding metric tensors,~\eqref{eq:metric_gauss1} and~\eqref{eq:metric_gauss2}, that are stated in terms of the Gaussian equilibrium covariance matrix and mean quadrature variables. Due to the non-commutativity between the conjugate forces acting on the system, it is clear that~\eqref{eq:metric_gauss1} and~\eqref{eq:metric_gauss2} are not typically proportional to each other, as we expect due to the breakdown of the work FDR indicated by~\eqref{eq:qFDR}. Minimising either the excess work or the work variance thus requires solving two different geodesic equations, which we illustrated with three examples. In the first example, we showed that taking a classical limit $\hbar\ll 1$ recovers the work FDR~\eqref{eq:FDR} and ensures the path of minimal average excess work coincides with the path of minimal work variance. When the dissipative dynamics has a single relaxation timescale, geodesic solutions can be found using results from information geometry for the multivariate normal distribution. In our second example, we found that the optimal way to displace an open Gaussian system with a constant covariance matrix is via a linear protocol even in a fully quantum regime. We found that this protocol can simultaneously optimise the average and variance in work despite the breakdown of the work FDR and quantum fluctuations in the conjugate variables. In the third example, we considered the optimisation of a damped quantum harmonic oscillator with control over the frequency and average position. We found that in this case, the geodesic for minimal excess work can substantially deviate from the geodesic minimising the fluctuations. Choosing either path was found to give significant improvements over a naive protocol, though following the path of minimal excess work was shown to lead to more dispersion in the overall work distribution that grows larger at smaller temperatures. This highlights the importance of considering the impact of quantum fluctuations and uncertainty for work processes in small scale systems. For small systems there is an inevitable compromise between the average work cost/gain along a process and the price one has to pay in increased fluctuations, and our method provides a geometric interpretation of this trade-off. More specifically, while the thermodynamic length may be shortened by a geodesic path with respect to one choice of metric, such as~\eqref{eq:metric_gauss1}, this may not be the shortest distance with respect to another metric such as~\eqref{eq:metric_gauss2} due to a difference in curvature. It is worth comparing these results to that of \cite{Miller2021}, which considered the problem of minimising the thermodynamic efficiency and work fluctuations in microscopic engines around a closed cycle. This represents a different kind of optimisation problem, as one fixes the path and needs to find the optimal speed at which to move around the cycle rather than determining a geodesic. In the case of speed optimisation, geometric trade-offs between efficiency and work fluctuations were found to hold, although it is worth noting that this trade-off is also applicable to classical heat engines. This is due to the fact that these quantities are not typically proportional to each other even classically. In contrast, the trade-offs between excess work and its variance that we observe in this paper arise solely from the fact that quantum fluctuations can be generated along a driving process, and this behaviour does not arise in classical systems close to equilibrium. If one were to drive a classical system further from equilibrium, then trade-offs between excess work and its variance begin to occur and the work FDR can be violated regardless of the presence of quantum fluctuations \cite{Solon2018a}. In light of this, it would be interesting to extend our results to Gaussian processes that move far from equilibrium, beyond the slow driving approximation, and understand how quantum effects influence this trade-off. Excess work minimisation beyond the slow driving regime has been achieved in a number of classical-mechanical examples involving driven Gaussian states \cite{Schmiedl,Blaber2021}, so one might hope that similar progress can be made in the quantum regime.

\begin{acknowledgments}
M. M. acknowledges  financial support from the Swiss National Science Foundation (NCCR SwissMAP).  H. J. D. M. acknowledges support from the Royal Commission for the Exhibition of 1851.
\end{acknowledgments}

\bibliographystyle{apsrev4-1}
\bibliography{mybib.bib}

\begin{thebibliography}{64}%
\makeatletter
\providecommand \@ifxundefined [1]{%
 \@ifx{#1\undefined}
}%
\providecommand \@ifnum [1]{%
 \ifnum #1\expandafter \@firstoftwo
 \else \expandafter \@secondoftwo
 \fi
}%
\providecommand \@ifx [1]{%
 \ifx #1\expandafter \@firstoftwo
 \else \expandafter \@secondoftwo
 \fi
}%
\providecommand \natexlab [1]{#1}%
\providecommand \enquote  [1]{``#1''}%
\providecommand \bibnamefont  [1]{#1}%
\providecommand \bibfnamefont [1]{#1}%
\providecommand \citenamefont [1]{#1}%
\providecommand \href@noop [0]{\@secondoftwo}%
\providecommand \href [0]{\begingroup \@sanitize@url \@href}%
\providecommand \@href[1]{\@@startlink{#1}\@@href}%
\providecommand \@@href[1]{\endgroup#1\@@endlink}%
\providecommand \@sanitize@url [0]{\catcode `\\12\catcode `\$12\catcode
  `\&12\catcode `\#12\catcode `\^12\catcode `\_12\catcode `\%12\relax}%
\providecommand \@@startlink[1]{}%
\providecommand \@@endlink[0]{}%
\providecommand \url  [0]{\begingroup\@sanitize@url \@url }%
\providecommand \@url [1]{\endgroup\@href {#1}{\urlprefix }}%
\providecommand \urlprefix  [0]{URL }%
\providecommand \Eprint [0]{\href }%
\providecommand \doibase [0]{http://dx.doi.org/}%
\providecommand \selectlanguage [0]{\@gobble}%
\providecommand \bibinfo  [0]{\@secondoftwo}%
\providecommand \bibfield  [0]{\@secondoftwo}%
\providecommand \translation [1]{[#1]}%
\providecommand \BibitemOpen [0]{}%
\providecommand \bibitemStop [0]{}%
\providecommand \bibitemNoStop [0]{.\EOS\space}%
\providecommand \EOS [0]{\spacefactor3000\relax}%
\providecommand \BibitemShut  [1]{\csname bibitem#1\endcsname}%
\let\auto@bib@innerbib\@empty
\bibitem [{\citenamefont {Seifert}(2012)}]{Seifert2012}%
  \BibitemOpen
  \bibfield  {author} {\bibinfo {author} {\bibfnamefont {U.}~\bibnamefont
  {Seifert}},\ }\href@noop {} {\bibfield  {journal} {\bibinfo  {journal} {Rep.
  Prog. Phys}\ }\textbf {\bibinfo {volume} {75}},\ \bibinfo {pages} {126001}
  (\bibinfo {year} {2012})}\BibitemShut {NoStop}%
\bibitem [{\citenamefont {Jarzynski}(1997)}]{Jarzynski1997c}%
  \BibitemOpen
  \bibfield  {author} {\bibinfo {author} {\bibfnamefont {C.}~\bibnamefont
  {Jarzynski}},\ }\href {\doibase 10.1103/PhysRevLett.78.2690} {\bibfield
  {journal} {\bibinfo  {journal} {Phys. Rev. Lett.}\ }\textbf {\bibinfo
  {volume} {78}},\ \bibinfo {pages} {2690} (\bibinfo {year}
  {1997})}\BibitemShut {NoStop}%
\bibitem [{\citenamefont {Crooks}(1999)}]{Crooks}%
  \BibitemOpen
  \bibfield  {author} {\bibinfo {author} {\bibfnamefont {G.}~\bibnamefont
  {Crooks}},\ }\href {\doibase 10.1103/PhysRevE.60.2721} {\bibfield  {journal}
  {\bibinfo  {journal} {Phys. Rev. E}\ }\textbf {\bibinfo {volume} {60}},\
  \bibinfo {pages} {2721} (\bibinfo {year} {1999})}\BibitemShut {NoStop}%
\bibitem [{\citenamefont {Campisi}\ \emph {et~al.}(2011)\citenamefont
  {Campisi}, \citenamefont {H{\"{a}}nggi},\ and\ \citenamefont
  {Talkner}}]{Campisi2011}%
  \BibitemOpen
  \bibfield  {author} {\bibinfo {author} {\bibfnamefont {M.}~\bibnamefont
  {Campisi}}, \bibinfo {author} {\bibfnamefont {P.}~\bibnamefont
  {H{\"{a}}nggi}}, \ and\ \bibinfo {author} {\bibfnamefont {P.}~\bibnamefont
  {Talkner}},\ }\href {http://rmp.aps.org/abstract/RMP/v83/i3/p771_1}
  {\bibfield  {journal} {\bibinfo  {journal} {Rev. Mod. Phys.}\ }\textbf
  {\bibinfo {volume} {83}},\ \bibinfo {pages} {771} (\bibinfo {year}
  {2011})}\BibitemShut {NoStop}%
\bibitem [{\citenamefont {Vinjanampathy}\ and\ \citenamefont
  {Anders}(2016)}]{Vinjanampathy2015b}%
  \BibitemOpen
  \bibfield  {author} {\bibinfo {author} {\bibfnamefont {S.}~\bibnamefont
  {Vinjanampathy}}\ and\ \bibinfo {author} {\bibfnamefont {J.}~\bibnamefont
  {Anders}},\ }\href {\doibase 10.1080/00107514.2016.1201896} {\bibfield
  {journal} {\bibinfo  {journal} {Cont. Phys.}\ }\textbf {\bibinfo {volume}
  {57}},\ \bibinfo {pages} {545} (\bibinfo {year} {2016})}\BibitemShut
  {NoStop}%
\bibitem [{\citenamefont {Allahverdyan}(2014)}]{Allahverdyan2014c}%
  \BibitemOpen
  \bibfield  {author} {\bibinfo {author} {\bibfnamefont {A.~E.}\ \bibnamefont
  {Allahverdyan}},\ }\href {\doibase 10.1103/PhysRevE.90.032137} {\bibfield
  {journal} {\bibinfo  {journal} {Phys. Rev. E}\ }\textbf {\bibinfo {volume}
  {90}},\ \bibinfo {pages} {032137} (\bibinfo {year} {2014})}\BibitemShut
  {NoStop}%
\bibitem [{\citenamefont {B{\"{a}}umer}\ \emph {et~al.}(2018)\citenamefont
  {B{\"{a}}umer}, \citenamefont {Lostaglio}, \citenamefont {Perarnau-Llobet},\
  and\ \citenamefont {Sampaio}}]{Baumer2018}%
  \BibitemOpen
  \bibfield  {author} {\bibinfo {author} {\bibfnamefont {E.}~\bibnamefont
  {B{\"{a}}umer}}, \bibinfo {author} {\bibfnamefont {M.}~\bibnamefont
  {Lostaglio}}, \bibinfo {author} {\bibfnamefont {M.}~\bibnamefont
  {Perarnau-Llobet}}, \ and\ \bibinfo {author} {\bibfnamefont {R.}~\bibnamefont
  {Sampaio}},\ }\enquote {\bibinfo {title} {{Fluctuating Work in Coherent
  Quantum Systems: Proposals and Limitations}},}\ in\ \href {\doibase
  10.1007/978-3-319-99046-0_11} {\emph {\bibinfo {booktitle} {Thermodynamics in
  the Quantum Regime: Fundamental Aspects and New Directions}}},\ \bibinfo
  {editor} {edited by\ \bibinfo {editor} {\bibfnamefont {F.}~\bibnamefont
  {Binder}}, \bibinfo {editor} {\bibfnamefont {L.~A.}\ \bibnamefont {Correa}},
  \bibinfo {editor} {\bibfnamefont {C.}~\bibnamefont {Gogolin}}, \bibinfo
  {editor} {\bibfnamefont {J.}~\bibnamefont {Anders}}, \ and\ \bibinfo {editor}
  {\bibfnamefont {G.}~\bibnamefont {Adesso}}}\ (\bibinfo  {publisher} {Springer
  International Publishing},\ \bibinfo {address} {Cham},\ \bibinfo {year}
  {2018})\ pp.\ \bibinfo {pages} {275--300}\BibitemShut {NoStop}%
\bibitem [{\citenamefont {Strasberg}(2019)}]{Strasberg2019}%
  \BibitemOpen
  \bibfield  {author} {\bibinfo {author} {\bibfnamefont {P.}~\bibnamefont
  {Strasberg}},\ }\href {\doibase 10.1103/PhysRevE.100.022127} {\bibfield
  {journal} {\bibinfo  {journal} {Phys. Rev. E}\ }\textbf {\bibinfo {volume}
  {100}},\ \bibinfo {pages} {022127} (\bibinfo {year} {2019})}\BibitemShut
  {NoStop}%
\bibitem [{\citenamefont {Perarnau-Llobet}\ \emph {et~al.}(2015)\citenamefont
  {Perarnau-Llobet}, \citenamefont {Hovhannisyan}, \citenamefont {Huber},
  \citenamefont {Skrzypczyk}, \citenamefont {Brunner},\ and\ \citenamefont
  {Ac{\'{i}}n}}]{Perarnau-Llobet2015b}%
  \BibitemOpen
  \bibfield  {author} {\bibinfo {author} {\bibfnamefont {M.}~\bibnamefont
  {Perarnau-Llobet}}, \bibinfo {author} {\bibfnamefont {K.~V.}\ \bibnamefont
  {Hovhannisyan}}, \bibinfo {author} {\bibfnamefont {M.}~\bibnamefont {Huber}},
  \bibinfo {author} {\bibfnamefont {P.}~\bibnamefont {Skrzypczyk}}, \bibinfo
  {author} {\bibfnamefont {N.}~\bibnamefont {Brunner}}, \ and\ \bibinfo
  {author} {\bibfnamefont {A.}~\bibnamefont {Ac{\'{i}}n}},\ }\href {\doibase
  10.1103/PhysRevX.5.041011} {\bibfield  {journal} {\bibinfo  {journal} {Phys.
  Rev. X}\ }\textbf {\bibinfo {volume} {5}},\ \bibinfo {pages} {041011}
  (\bibinfo {year} {2015})}\BibitemShut {NoStop}%
\bibitem [{\citenamefont {Solinas}\ and\ \citenamefont
  {Gasparinetti}(2015)}]{Solinas2018}%
  \BibitemOpen
  \bibfield  {author} {\bibinfo {author} {\bibfnamefont {P.}~\bibnamefont
  {Solinas}}\ and\ \bibinfo {author} {\bibfnamefont {S.}~\bibnamefont
  {Gasparinetti}},\ }\href@noop {} {\bibfield  {journal} {\bibinfo  {journal}
  {Phys. Rev. E}\ }\textbf {\bibinfo {volume} {92}},\ \bibinfo {pages} {042150}
  (\bibinfo {year} {2015})}\BibitemShut {NoStop}%
\bibitem [{\citenamefont {Miller}\ and\ \citenamefont
  {Anders}(2018)}]{Miller2018a}%
  \BibitemOpen
  \bibfield  {author} {\bibinfo {author} {\bibfnamefont {H.~J.~D.}\
  \bibnamefont {Miller}}\ and\ \bibinfo {author} {\bibfnamefont
  {J.}~\bibnamefont {Anders}},\ }\href@noop {} {\bibfield  {journal} {\bibinfo
  {journal} {Entropy}\ }\textbf {\bibinfo {volume} {20}},\ \bibinfo {pages}
  {200} (\bibinfo {year} {2018})}\BibitemShut {NoStop}%
\bibitem [{\citenamefont {Lostaglio}(2018)}]{Lostaglio2018}%
  \BibitemOpen
  \bibfield  {author} {\bibinfo {author} {\bibfnamefont {M.}~\bibnamefont
  {Lostaglio}},\ }\href {\doibase 10.1103/PhysRevLett.120.040602} {\bibfield
  {journal} {\bibinfo  {journal} {Phys. Rev. Lett.}\ }\textbf {\bibinfo
  {volume} {120}},\ \bibinfo {pages} {040602} (\bibinfo {year}
  {2018})}\BibitemShut {NoStop}%
\bibitem [{\citenamefont {Korzekwa}\ \emph {et~al.}(2016)\citenamefont
  {Korzekwa}, \citenamefont {Lostaglio}, \citenamefont {Oppenheim},\ and\
  \citenamefont {Jennings}}]{Korzekwa2016}%
  \BibitemOpen
  \bibfield  {author} {\bibinfo {author} {\bibfnamefont {K.}~\bibnamefont
  {Korzekwa}}, \bibinfo {author} {\bibfnamefont {M.}~\bibnamefont {Lostaglio}},
  \bibinfo {author} {\bibfnamefont {J.}~\bibnamefont {Oppenheim}}, \ and\
  \bibinfo {author} {\bibfnamefont {D.}~\bibnamefont {Jennings}},\ }\href
  {\doibase 10.1088/1367-2630/18/2/023045} {\bibfield  {journal} {\bibinfo
  {journal} {N. J. Phys}\ }\textbf {\bibinfo {volume} {18}},\ \bibinfo {pages}
  {023045} (\bibinfo {year} {2016})}\BibitemShut {NoStop}%
\bibitem [{\citenamefont {Brandner}\ \emph {et~al.}(2017)\citenamefont
  {Brandner}, \citenamefont {Bauer},\ and\ \citenamefont {Seifert}}]{Brandner}%
  \BibitemOpen
  \bibfield  {author} {\bibinfo {author} {\bibfnamefont {K.}~\bibnamefont
  {Brandner}}, \bibinfo {author} {\bibfnamefont {M.}~\bibnamefont {Bauer}}, \
  and\ \bibinfo {author} {\bibfnamefont {U.}~\bibnamefont {Seifert}},\
  }\href@noop {} {\bibfield  {journal} {\bibinfo  {journal} {Phys. Rev. Lett.}\
  }\textbf {\bibinfo {volume} {119}},\ \bibinfo {pages} {170602} (\bibinfo
  {year} {2017})}\BibitemShut {NoStop}%
\bibitem [{\citenamefont {Varizi}\ \emph {et~al.}(2021)\citenamefont {Varizi},
  \citenamefont {Cipolla}, \citenamefont {Perarnau-Llobet}, \citenamefont
  {Drumond},\ and\ \citenamefont {Landi}}]{Varizi2021}%
  \BibitemOpen
  \bibfield  {author} {\bibinfo {author} {\bibfnamefont {A.~D.}\ \bibnamefont
  {Varizi}}, \bibinfo {author} {\bibfnamefont {M.~A.}\ \bibnamefont {Cipolla}},
  \bibinfo {author} {\bibfnamefont {M.}~\bibnamefont {Perarnau-Llobet}},
  \bibinfo {author} {\bibfnamefont {R.~C.}\ \bibnamefont {Drumond}}, \ and\
  \bibinfo {author} {\bibfnamefont {G.~T.}\ \bibnamefont {Landi}},\ }\href
  {\doibase 10.1088/1367-2630/abfe20} {\bibfield  {journal} {\bibinfo
  {journal} {N. J. Phys}\ }\textbf {\bibinfo {volume} {23}},\ \bibinfo {pages}
  {1} (\bibinfo {year} {2021})}\BibitemShut {NoStop}%
\bibitem [{\citenamefont {Francica}\ \emph {et~al.}(2020)\citenamefont
  {Francica}, \citenamefont {Binder}, \citenamefont {Guarnieri}, \citenamefont
  {Mitchison}, \citenamefont {Goold},\ and\ \citenamefont
  {Plastina}}]{Francica2020}%
  \BibitemOpen
  \bibfield  {author} {\bibinfo {author} {\bibfnamefont {G.}~\bibnamefont
  {Francica}}, \bibinfo {author} {\bibfnamefont {F.~C.}\ \bibnamefont
  {Binder}}, \bibinfo {author} {\bibfnamefont {G.}~\bibnamefont {Guarnieri}},
  \bibinfo {author} {\bibfnamefont {M.~T.}\ \bibnamefont {Mitchison}}, \bibinfo
  {author} {\bibfnamefont {J.}~\bibnamefont {Goold}}, \ and\ \bibinfo {author}
  {\bibfnamefont {F.}~\bibnamefont {Plastina}},\ }\href@noop {} {\bibfield
  {journal} {\bibinfo  {journal} {Phys. Rev. Lett.}\ }\textbf {\bibinfo
  {volume} {125}},\ \bibinfo {pages} {180603} (\bibinfo {year}
  {2020})}\BibitemShut {NoStop}%
\bibitem [{\citenamefont {Mohammady}\ and\ \citenamefont
  {Romito}(2019)}]{Mohammady2019}%
  \BibitemOpen
  \bibfield  {author} {\bibinfo {author} {\bibfnamefont {M.~H.}\ \bibnamefont
  {Mohammady}}\ and\ \bibinfo {author} {\bibfnamefont {A.}~\bibnamefont
  {Romito}},\ }\href {\doibase 10.22331/q-2019-08-19-175} {\bibfield  {journal}
  {\bibinfo  {journal} {Quantum}\ }\textbf {\bibinfo {volume} {3}},\ \bibinfo
  {pages} {175} (\bibinfo {year} {2019})}\BibitemShut {NoStop}%
\bibitem [{\citenamefont {Speck}\ and\ \citenamefont {Seifert}(2004)}]{Speck}%
  \BibitemOpen
  \bibfield  {author} {\bibinfo {author} {\bibfnamefont {T.}~\bibnamefont
  {Speck}}\ and\ \bibinfo {author} {\bibfnamefont {U.}~\bibnamefont
  {Seifert}},\ }\href@noop {} {\bibfield  {journal} {\bibinfo  {journal} {Phys.
  Rev. E}\ }\textbf {\bibinfo {volume} {70}},\ \bibinfo {pages} {066112}
  (\bibinfo {year} {2004})}\BibitemShut {NoStop}%
\bibitem [{\citenamefont {Mandal}\ and\ \citenamefont
  {Jarzynski}(2016)}]{Mandal2016a}%
  \BibitemOpen
  \bibfield  {author} {\bibinfo {author} {\bibfnamefont {D.}~\bibnamefont
  {Mandal}}\ and\ \bibinfo {author} {\bibfnamefont {C.}~\bibnamefont
  {Jarzynski}},\ }\href {\doibase 10.1088/1742-5468/2016/06/063204} {\bibfield
  {journal} {\bibinfo  {journal} {J. Stat. Mech.}\ }\textbf {\bibinfo {volume}
  {2016}},\ \bibinfo {pages} {063204} (\bibinfo {year} {2016})}\BibitemShut
  {NoStop}%
\bibitem [{\citenamefont {Miller}\ \emph {et~al.}(2019)\citenamefont {Miller},
  \citenamefont {Scandi}, \citenamefont {Anders},\ and\ \citenamefont
  {Perarnau-Llobet}}]{Miller2019}%
  \BibitemOpen
  \bibfield  {author} {\bibinfo {author} {\bibfnamefont {H.~J.~D.}\
  \bibnamefont {Miller}}, \bibinfo {author} {\bibfnamefont {M.}~\bibnamefont
  {Scandi}}, \bibinfo {author} {\bibfnamefont {J.}~\bibnamefont {Anders}}, \
  and\ \bibinfo {author} {\bibfnamefont {M.}~\bibnamefont {Perarnau-Llobet}},\
  }\href@noop {} {\bibfield  {journal} {\bibinfo  {journal} {Phys. Rev. Lett.}\
  }\textbf {\bibinfo {volume} {123}},\ \bibinfo {pages} {230603} (\bibinfo
  {year} {2019})}\BibitemShut {NoStop}%
\bibitem [{\citenamefont {Scandi}\ \emph {et~al.}(2020)\citenamefont {Scandi},
  \citenamefont {Miller}, \citenamefont {Anders},\ and\ \citenamefont
  {Perarnau-Llobet}}]{Scandi2019}%
  \BibitemOpen
  \bibfield  {author} {\bibinfo {author} {\bibfnamefont {M.}~\bibnamefont
  {Scandi}}, \bibinfo {author} {\bibfnamefont {H.~J.~D.}\ \bibnamefont
  {Miller}}, \bibinfo {author} {\bibfnamefont {J.}~\bibnamefont {Anders}}, \
  and\ \bibinfo {author} {\bibfnamefont {M.}~\bibnamefont {Perarnau-Llobet}},\
  }\href@noop {} {\bibfield  {journal} {\bibinfo  {journal} {Phys. Rev.
  Research}\ }\textbf {\bibinfo {volume} {2}},\ \bibinfo {pages} {023377}
  (\bibinfo {year} {2020})}\BibitemShut {NoStop}%
\bibitem [{\citenamefont {Wigner}\ and\ \citenamefont
  {Yanase}(1963)}]{Wigner1963a}%
  \BibitemOpen
  \bibfield  {author} {\bibinfo {author} {\bibfnamefont {E.~P.}\ \bibnamefont
  {Wigner}}\ and\ \bibinfo {author} {\bibfnamefont {M.~M.}\ \bibnamefont
  {Yanase}},\ }\href {\doibase 10.1073/pnas.49.6.910} {\bibfield  {journal}
  {\bibinfo  {journal} {J. Phys. Chem.}\ }\textbf {\bibinfo {volume} {15}},\
  \bibinfo {pages} {1084} (\bibinfo {year} {1963})}\BibitemShut {NoStop}%
\bibitem [{\citenamefont {Luo}(2006)}]{Luo2006}%
  \BibitemOpen
  \bibfield  {author} {\bibinfo {author} {\bibfnamefont {S.}~\bibnamefont
  {Luo}},\ }\href {\doibase 10.1103/PhysRevA.73.022324} {\bibfield  {journal}
  {\bibinfo  {journal} {Phys. Rev. A}\ }\textbf {\bibinfo {volume} {73}},\
  \bibinfo {pages} {022324} (\bibinfo {year} {2006})}\BibitemShut {NoStop}%
\bibitem [{\citenamefont {Ruppeiner}(1979)}]{Ruppeiner1979}%
  \BibitemOpen
  \bibfield  {author} {\bibinfo {author} {\bibfnamefont {G.}~\bibnamefont
  {Ruppeiner}},\ }\href {\doibase 10.1103/PhysRevA.20.1608} {\bibfield
  {journal} {\bibinfo  {journal} {Phys. Rev. A}\ }\textbf {\bibinfo {volume}
  {20}},\ \bibinfo {pages} {1608} (\bibinfo {year} {1979})}\BibitemShut
  {NoStop}%
\bibitem [{\citenamefont {Salamon}\ and\ \citenamefont
  {Berry}(1983)}]{Salamon1983}%
  \BibitemOpen
  \bibfield  {author} {\bibinfo {author} {\bibfnamefont {P.}~\bibnamefont
  {Salamon}}\ and\ \bibinfo {author} {\bibfnamefont {R.~S.}\ \bibnamefont
  {Berry}},\ }\href {\doibase 10.1103/PhysRevLett.51.1127} {\bibfield
  {journal} {\bibinfo  {journal} {Phys. Rev. Lett.}\ }\textbf {\bibinfo
  {volume} {51}},\ \bibinfo {pages} {1127} (\bibinfo {year}
  {1983})}\BibitemShut {NoStop}%
\bibitem [{\citenamefont {Nulton}\ \emph {et~al.}(1985)\citenamefont {Nulton},
  \citenamefont {Salamon}, \citenamefont {Andresen}, \citenamefont {Anmin},
  \citenamefont {Nulton}, \citenamefont {Salamon}, \citenamefont {Andresen},\
  and\ \citenamefont {Anmin}}]{Nulton2013}%
  \BibitemOpen
  \bibfield  {author} {\bibinfo {author} {\bibfnamefont {J.}~\bibnamefont
  {Nulton}}, \bibinfo {author} {\bibfnamefont {P.}~\bibnamefont {Salamon}},
  \bibinfo {author} {\bibfnamefont {B.}~\bibnamefont {Andresen}}, \bibinfo
  {author} {\bibfnamefont {Q.}~\bibnamefont {Anmin}}, \bibinfo {author}
  {\bibfnamefont {J.}~\bibnamefont {Nulton}}, \bibinfo {author} {\bibfnamefont
  {P.}~\bibnamefont {Salamon}}, \bibinfo {author} {\bibfnamefont
  {B.}~\bibnamefont {Andresen}}, \ and\ \bibinfo {author} {\bibfnamefont
  {Q.}~\bibnamefont {Anmin}},\ }\href {\doibase 10.1063/1.449774} {\bibfield
  {journal} {\bibinfo  {journal} {J. Chem. Phys.}\ }\textbf {\bibinfo {volume}
  {83}},\ \bibinfo {pages} {334} (\bibinfo {year} {1985})}\BibitemShut
  {NoStop}%
\bibitem [{\citenamefont {Crooks}(2007)}]{Crooks2007}%
  \BibitemOpen
  \bibfield  {author} {\bibinfo {author} {\bibfnamefont {G.~E.}\ \bibnamefont
  {Crooks}},\ }\href {\doibase 10.1103/PhysRevLett.99.100602} {\bibfield
  {journal} {\bibinfo  {journal} {Phys. Rev. Lett.}\ }\textbf {\bibinfo
  {volume} {99}},\ \bibinfo {pages} {100602} (\bibinfo {year}
  {2007})}\BibitemShut {NoStop}%
\bibitem [{\citenamefont {Sivak}\ and\ \citenamefont
  {Crooks}(2012)}]{Sivak2012a}%
  \BibitemOpen
  \bibfield  {author} {\bibinfo {author} {\bibfnamefont {D.~A.}\ \bibnamefont
  {Sivak}}\ and\ \bibinfo {author} {\bibfnamefont {G.~E.}\ \bibnamefont
  {Crooks}},\ }\href {\doibase 10.1103/PhysRevLett.108.190602} {\bibfield
  {journal} {\bibinfo  {journal} {Phys. Rev. Lett.}\ }\textbf {\bibinfo
  {volume} {108}},\ \bibinfo {pages} {190602} (\bibinfo {year}
  {2012})}\BibitemShut {NoStop}%
\bibitem [{\citenamefont {Zulkowski}\ \emph {et~al.}(2012)\citenamefont
  {Zulkowski}, \citenamefont {Sivak}, \citenamefont {Crooks},\ and\
  \citenamefont {Deweese}}]{Zulkowski2012}%
  \BibitemOpen
  \bibfield  {author} {\bibinfo {author} {\bibfnamefont {P.~R.}\ \bibnamefont
  {Zulkowski}}, \bibinfo {author} {\bibfnamefont {D.~A.}\ \bibnamefont
  {Sivak}}, \bibinfo {author} {\bibfnamefont {G.~E.}\ \bibnamefont {Crooks}}, \
  and\ \bibinfo {author} {\bibfnamefont {M.~R.}\ \bibnamefont {Deweese}},\
  }\href {\doibase 10.1103/PhysRevE.86.041148} {\bibfield  {journal} {\bibinfo
  {journal} {Phys. Rev. E}\ }\textbf {\bibinfo {volume} {86}},\ \bibinfo
  {pages} {0141148} (\bibinfo {year} {2012})}\BibitemShut {NoStop}%
\bibitem [{\citenamefont {Abiuso}\ \emph {et~al.}(2020)\citenamefont {Abiuso},
  \citenamefont {Miller}, \citenamefont {Perarnau-Llobet},\ and\ \citenamefont
  {Scandi}}]{Abiuso2020a}%
  \BibitemOpen
  \bibfield  {author} {\bibinfo {author} {\bibfnamefont {P.}~\bibnamefont
  {Abiuso}}, \bibinfo {author} {\bibfnamefont {H.~J.}\ \bibnamefont {Miller}},
  \bibinfo {author} {\bibfnamefont {M.}~\bibnamefont {Perarnau-Llobet}}, \ and\
  \bibinfo {author} {\bibfnamefont {M.}~\bibnamefont {Scandi}},\ }\href
  {\doibase 10.3390/e22101076} {\bibfield  {journal} {\bibinfo  {journal}
  {Entropy}\ }\textbf {\bibinfo {volume} {22}},\ \bibinfo {pages} {1076}
  (\bibinfo {year} {2020})}\BibitemShut {NoStop}%
\bibitem [{\citenamefont {Burbea}(1986)}]{Burbea}%
  \BibitemOpen
  \bibfield  {author} {\bibinfo {author} {\bibfnamefont {J.}~\bibnamefont
  {Burbea}},\ }\href@noop {} {\bibfield  {journal} {\bibinfo  {journal} {Expo.
  Math.}\ }\textbf {\bibinfo {volume} {4}},\ \bibinfo {pages} {347} (\bibinfo
  {year} {1986})}\BibitemShut {NoStop}%
\bibitem [{\citenamefont {Amari}\ and\ \citenamefont
  {Nagaoka}(2007)}]{Amari2007}%
  \BibitemOpen
  \bibfield  {author} {\bibinfo {author} {\bibfnamefont {S.~I.}\ \bibnamefont
  {Amari}}\ and\ \bibinfo {author} {\bibfnamefont {H.}~\bibnamefont
  {Nagaoka}},\ }\href@noop {} {\emph {\bibinfo {title} {{Methods of information
  geometry}}}}\ (\bibinfo  {publisher} {American Mathematical Soc.},\ \bibinfo
  {year} {2007})\BibitemShut {NoStop}%
\bibitem [{\citenamefont {Miller}\ and\ \citenamefont
  {Mehboudi}(2020)}]{Miller2020}%
  \BibitemOpen
  \bibfield  {author} {\bibinfo {author} {\bibfnamefont {H.~J.}\ \bibnamefont
  {Miller}}\ and\ \bibinfo {author} {\bibfnamefont {M.}~\bibnamefont
  {Mehboudi}},\ }\href {\doibase 10.1103/PhysRevLett.125.260602} {\bibfield
  {journal} {\bibinfo  {journal} {Phys. Rev. Lett.}\ }\textbf {\bibinfo
  {volume} {125}},\ \bibinfo {pages} {260602} (\bibinfo {year}
  {2020})}\BibitemShut {NoStop}%
\bibitem [{\citenamefont {Weedbrook}\ \emph {et~al.}(2012)\citenamefont
  {Weedbrook}, \citenamefont {Pirandola}, \citenamefont {Cerf},\ and\
  \citenamefont {Ralph}}]{Weedbrook}%
  \BibitemOpen
  \bibfield  {author} {\bibinfo {author} {\bibfnamefont {C.}~\bibnamefont
  {Weedbrook}}, \bibinfo {author} {\bibfnamefont {S.}~\bibnamefont
  {Pirandola}}, \bibinfo {author} {\bibfnamefont {N.~J.}\ \bibnamefont {Cerf}},
  \ and\ \bibinfo {author} {\bibfnamefont {T.~C.}\ \bibnamefont {Ralph}},\
  }\href@noop {} {\bibfield  {journal} {\bibinfo  {journal} {Rev. Mod. Phys.}\
  }\textbf {\bibinfo {volume} {84}},\ \bibinfo {pages} {621} (\bibinfo {year}
  {2012})}\BibitemShut {NoStop}%
\bibitem [{\citenamefont {Adesso}\ \emph {et~al.}(2014)\citenamefont {Adesso},
  \citenamefont {Ragy},\ and\ \citenamefont {Lee}}]{Adesso2014a}%
  \BibitemOpen
  \bibfield  {author} {\bibinfo {author} {\bibfnamefont {G.}~\bibnamefont
  {Adesso}}, \bibinfo {author} {\bibfnamefont {S.}~\bibnamefont {Ragy}}, \ and\
  \bibinfo {author} {\bibfnamefont {A.~R.}\ \bibnamefont {Lee}},\ }\href
  {\doibase 10.1142/S1230161214400010} {\bibfield  {journal} {\bibinfo
  {journal} {Open Sys. and Info. Dyn.}\ }\textbf {\bibinfo {volume} {21}},\
  \bibinfo {pages} {1} (\bibinfo {year} {2014})}\BibitemShut {NoStop}%
\bibitem [{\citenamefont {Mehboudi}\ and\ \citenamefont
  {Parrondo}(2019)}]{Mehboudi}%
  \BibitemOpen
  \bibfield  {author} {\bibinfo {author} {\bibfnamefont {M.}~\bibnamefont
  {Mehboudi}}\ and\ \bibinfo {author} {\bibfnamefont {J.~M.~R.}\ \bibnamefont
  {Parrondo}},\ }\href@noop {} {\bibfield  {journal} {\bibinfo  {journal} {N.
  J. Phys}\ }\textbf {\bibinfo {volume} {21}},\ \bibinfo {pages} {083036}
  (\bibinfo {year} {2019})}\BibitemShut {NoStop}%
\bibitem [{\citenamefont {Brown}\ \emph {et~al.}(2016)\citenamefont {Brown},
  \citenamefont {Friis},\ and\ \citenamefont {Huber}}]{Brown2016c}%
  \BibitemOpen
  \bibfield  {author} {\bibinfo {author} {\bibfnamefont {E.~G.}\ \bibnamefont
  {Brown}}, \bibinfo {author} {\bibfnamefont {N.}~\bibnamefont {Friis}}, \ and\
  \bibinfo {author} {\bibfnamefont {M.}~\bibnamefont {Huber}},\ }\href
  {\doibase 10.1088/1367-2630/18/11/113028} {\bibfield  {journal} {\bibinfo
  {journal} {N. J. Phys}\ }\textbf {\bibinfo {volume} {18}},\ \bibinfo {pages}
  {113028} (\bibinfo {year} {2016})}\BibitemShut {NoStop}%
\bibitem [{\citenamefont {Singh}\ \emph {et~al.}(2019)\citenamefont {Singh},
  \citenamefont {Jabbour}, \citenamefont {{Van Herstraeten}},\ and\
  \citenamefont {Cerf}}]{Singh2019}%
  \BibitemOpen
  \bibfield  {author} {\bibinfo {author} {\bibfnamefont {U.}~\bibnamefont
  {Singh}}, \bibinfo {author} {\bibfnamefont {M.~G.}\ \bibnamefont {Jabbour}},
  \bibinfo {author} {\bibfnamefont {Z.}~\bibnamefont {{Van Herstraeten}}}, \
  and\ \bibinfo {author} {\bibfnamefont {N.~J.}\ \bibnamefont {Cerf}},\ }\href
  {\doibase 10.1103/PhysRevA.100.042104} {\bibfield  {journal} {\bibinfo
  {journal} {Phys. Rev. A}\ }\textbf {\bibinfo {volume} {100}},\ \bibinfo
  {pages} {042104} (\bibinfo {year} {2019})}\BibitemShut {NoStop}%
\bibitem [{\citenamefont {Belenchia}\ \emph {et~al.}(2020)\citenamefont
  {Belenchia}, \citenamefont {Mancino}, \citenamefont {Landi},\ and\
  \citenamefont {Paternostro}}]{Belenchia2020a}%
  \BibitemOpen
  \bibfield  {author} {\bibinfo {author} {\bibfnamefont {A.}~\bibnamefont
  {Belenchia}}, \bibinfo {author} {\bibfnamefont {L.}~\bibnamefont {Mancino}},
  \bibinfo {author} {\bibfnamefont {G.~T.}\ \bibnamefont {Landi}}, \ and\
  \bibinfo {author} {\bibfnamefont {M.}~\bibnamefont {Paternostro}},\ }\href
  {http://dx.doi.org/10.1038/s41534-020-00334-6} {\bibfield  {journal}
  {\bibinfo  {journal} {npj Quant. Info.}\ }\textbf {\bibinfo {volume} {6}}
  (\bibinfo {year} {2020})}\BibitemShut {NoStop}%
\bibitem [{\citenamefont {Zanin}\ \emph {et~al.}(2019)\citenamefont {Zanin},
  \citenamefont {H{\"{a}}ffner}, \citenamefont {Talarico}, \citenamefont
  {Duzzioni}, \citenamefont {Ribeiro}, \citenamefont {Landi},\ and\
  \citenamefont {C{\'{e}}leri}}]{Zanin2019}%
  \BibitemOpen
  \bibfield  {author} {\bibinfo {author} {\bibfnamefont {G.~L.}\ \bibnamefont
  {Zanin}}, \bibinfo {author} {\bibfnamefont {T.}~\bibnamefont
  {H{\"{a}}ffner}}, \bibinfo {author} {\bibfnamefont {M.~A.}\ \bibnamefont
  {Talarico}}, \bibinfo {author} {\bibfnamefont {E.~I.}\ \bibnamefont
  {Duzzioni}}, \bibinfo {author} {\bibfnamefont {P.~H.}\ \bibnamefont
  {Ribeiro}}, \bibinfo {author} {\bibfnamefont {G.~T.}\ \bibnamefont {Landi}},
  \ and\ \bibinfo {author} {\bibfnamefont {L.~C.}\ \bibnamefont
  {C{\'{e}}leri}},\ }\href {\doibase 10.1007/s13538-019-00700-6} {\bibfield
  {journal} {\bibinfo  {journal} {Braz. J. Phys}\ }\textbf {\bibinfo {volume}
  {49}},\ \bibinfo {pages} {783} (\bibinfo {year} {2019})}\BibitemShut
  {NoStop}%
\bibitem [{\citenamefont {Albash}\ \emph {et~al.}(2012)\citenamefont {Albash},
  \citenamefont {Boixo},\ and\ \citenamefont {Lidar}}]{Albash2012}%
  \BibitemOpen
  \bibfield  {author} {\bibinfo {author} {\bibfnamefont {T.}~\bibnamefont
  {Albash}}, \bibinfo {author} {\bibfnamefont {S.}~\bibnamefont {Boixo}}, \
  and\ \bibinfo {author} {\bibfnamefont {D.~A.}\ \bibnamefont {Lidar}},\
  }\href@noop {} {\bibfield  {journal} {\bibinfo  {journal} {N. J. Phys}\
  }\textbf {\bibinfo {volume} {14}},\ \bibinfo {pages} {123016} (\bibinfo
  {year} {2012})}\BibitemShut {NoStop}%
\bibitem [{\citenamefont {Talkner}\ \emph {et~al.}(2007)\citenamefont
  {Talkner}, \citenamefont {Lutz},\ and\ \citenamefont
  {H{\"{a}}nggi}}]{Talkner2007c}%
  \BibitemOpen
  \bibfield  {author} {\bibinfo {author} {\bibfnamefont {P.}~\bibnamefont
  {Talkner}}, \bibinfo {author} {\bibfnamefont {E.}~\bibnamefont {Lutz}}, \
  and\ \bibinfo {author} {\bibfnamefont {P.}~\bibnamefont {H{\"{a}}nggi}},\
  }\href {\doibase 10.1103/PhysRevE.75.050102} {\bibfield  {journal} {\bibinfo
  {journal} {Phys. Rev. E}\ }\textbf {\bibinfo {volume} {75}},\ \bibinfo
  {pages} {050102} (\bibinfo {year} {2007})}\BibitemShut {NoStop}%
\bibitem [{\citenamefont {Esposito}\ \emph {et~al.}(2009)\citenamefont
  {Esposito}, \citenamefont {Harbola},\ and\ \citenamefont
  {Mukamel}}]{Esposito2009}%
  \BibitemOpen
  \bibfield  {author} {\bibinfo {author} {\bibfnamefont {M.}~\bibnamefont
  {Esposito}}, \bibinfo {author} {\bibfnamefont {U.}~\bibnamefont {Harbola}}, \
  and\ \bibinfo {author} {\bibfnamefont {S.}~\bibnamefont {Mukamel}},\ }\href
  {http://link.aps.org/doi/10.1103/RevModPhys.81.1665} {\bibfield  {journal}
  {\bibinfo  {journal} {Rev. Mod. Phys.}\ }\textbf {\bibinfo {volume} {81}},\
  \bibinfo {pages} {1665} (\bibinfo {year} {2009})}\BibitemShut {NoStop}%
\bibitem [{\citenamefont {Silaev}\ \emph {et~al.}(2014)\citenamefont {Silaev},
  \citenamefont {Heikkil{\"{a}}},\ and\ \citenamefont
  {Virtanen}}]{Silaev2014b}%
  \BibitemOpen
  \bibfield  {author} {\bibinfo {author} {\bibfnamefont {M.}~\bibnamefont
  {Silaev}}, \bibinfo {author} {\bibfnamefont {T.~T.}\ \bibnamefont
  {Heikkil{\"{a}}}}, \ and\ \bibinfo {author} {\bibfnamefont {P.}~\bibnamefont
  {Virtanen}},\ }\href {\doibase 10.1103/PhysRevE.90.022103} {\bibfield
  {journal} {\bibinfo  {journal} {Phys. Rev. E}\ }\textbf {\bibinfo {volume}
  {90}},\ \bibinfo {pages} {022103} (\bibinfo {year} {2014})}\BibitemShut
  {NoStop}%
\bibitem [{\citenamefont {Horowitz}(2012)}]{Horowitz2012}%
  \BibitemOpen
  \bibfield  {author} {\bibinfo {author} {\bibfnamefont {J.~M.}\ \bibnamefont
  {Horowitz}},\ }\href {\doibase 10.1103/PhysRevE.85.031110} {\bibfield
  {journal} {\bibinfo  {journal} {Phys. Rev. E}\ }\textbf {\bibinfo {volume}
  {85}},\ \bibinfo {pages} {031110} (\bibinfo {year} {2012})}\BibitemShut
  {NoStop}%
\bibitem [{\citenamefont {Horowitz}\ and\ \citenamefont
  {Parrondo}(2013)}]{Horowitz2013b}%
  \BibitemOpen
  \bibfield  {author} {\bibinfo {author} {\bibfnamefont {J.~M.}\ \bibnamefont
  {Horowitz}}\ and\ \bibinfo {author} {\bibfnamefont {J.~M.~R.}\ \bibnamefont
  {Parrondo}},\ }\href {\doibase 10.1088/1367-2630/15/8/085028} {\bibfield
  {journal} {\bibinfo  {journal} {N. J. Phys}\ }\textbf {\bibinfo {volume}
  {15}},\ \bibinfo {pages} {085028} (\bibinfo {year} {2013})}\BibitemShut
  {NoStop}%
\bibitem [{\citenamefont {Manzano}\ \emph {et~al.}(2015)\citenamefont
  {Manzano}, \citenamefont {Horowitz},\ and\ \citenamefont
  {Parrondo}}]{Manzano2015}%
  \BibitemOpen
  \bibfield  {author} {\bibinfo {author} {\bibfnamefont {G.}~\bibnamefont
  {Manzano}}, \bibinfo {author} {\bibfnamefont {J.~M.}\ \bibnamefont
  {Horowitz}}, \ and\ \bibinfo {author} {\bibfnamefont {J.~M.~R.}\ \bibnamefont
  {Parrondo}},\ }\href {\doibase 10.1103/PhysRevE.92.032129} {\bibfield
  {journal} {\bibinfo  {journal} {Phys. Rev. E}\ }\textbf {\bibinfo {volume}
  {92}},\ \bibinfo {pages} {032129} (\bibinfo {year} {2015})}\BibitemShut
  {NoStop}%
\bibitem [{\citenamefont {Liu}\ and\ \citenamefont {Xi}(2016)}]{Liu2016b}%
  \BibitemOpen
  \bibfield  {author} {\bibinfo {author} {\bibfnamefont {F.}~\bibnamefont
  {Liu}}\ and\ \bibinfo {author} {\bibfnamefont {J.}~\bibnamefont {Xi}},\
  }\href@noop {} {\bibfield  {journal} {\bibinfo  {journal} {Phys. Rev. E}\
  }\textbf {\bibinfo {volume} {94}},\ \bibinfo {pages} {062133} (\bibinfo
  {year} {2016})}\BibitemShut {NoStop}%
\bibitem [{\citenamefont {Miller}\ \emph {et~al.}(2021)\citenamefont {Miller},
  \citenamefont {Mohammady}, \citenamefont {Perarnau-Llobet},\ and\
  \citenamefont {Guarnieri}}]{Miller2021}%
  \BibitemOpen
  \bibfield  {author} {\bibinfo {author} {\bibfnamefont {H.~J.}\ \bibnamefont
  {Miller}}, \bibinfo {author} {\bibfnamefont {M.~H.}\ \bibnamefont
  {Mohammady}}, \bibinfo {author} {\bibfnamefont {M.}~\bibnamefont
  {Perarnau-Llobet}}, \ and\ \bibinfo {author} {\bibfnamefont {G.}~\bibnamefont
  {Guarnieri}},\ }\href {\doibase 10.1103/PhysRevE.103.052138} {\bibfield
  {journal} {\bibinfo  {journal} {Phys. Rev. E}\ }\textbf {\bibinfo {volume}
  {103}},\ \bibinfo {pages} {052138} (\bibinfo {year} {2021})}\BibitemShut
  {NoStop}%
\bibitem [{\citenamefont {Suomela}\ \emph {et~al.}(2014)\citenamefont
  {Suomela}, \citenamefont {Solinas}, \citenamefont {Pekola}, \citenamefont
  {Ankerhold},\ and\ \citenamefont {Ala-Nissila}}]{Suomela2014}%
  \BibitemOpen
  \bibfield  {author} {\bibinfo {author} {\bibfnamefont {S.}~\bibnamefont
  {Suomela}}, \bibinfo {author} {\bibfnamefont {P.}~\bibnamefont {Solinas}},
  \bibinfo {author} {\bibfnamefont {J.~P.}\ \bibnamefont {Pekola}}, \bibinfo
  {author} {\bibfnamefont {J.}~\bibnamefont {Ankerhold}}, \ and\ \bibinfo
  {author} {\bibfnamefont {T.}~\bibnamefont {Ala-Nissila}},\ }\href {\doibase
  10.1103/PhysRevB.90.094304} {\bibfield  {journal} {\bibinfo  {journal} {Phys.
  Rev. B}\ }\textbf {\bibinfo {volume} {90}},\ \bibinfo {pages} {094304}
  (\bibinfo {year} {2014})}\BibitemShut {NoStop}%
\bibitem [{\citenamefont {Cavina}\ \emph {et~al.}(2017)\citenamefont {Cavina},
  \citenamefont {Mari},\ and\ \citenamefont {Giovannetti}}]{Cavina2017}%
  \BibitemOpen
  \bibfield  {author} {\bibinfo {author} {\bibfnamefont {V.}~\bibnamefont
  {Cavina}}, \bibinfo {author} {\bibfnamefont {A.}~\bibnamefont {Mari}}, \ and\
  \bibinfo {author} {\bibfnamefont {V.}~\bibnamefont {Giovannetti}},\ }\href
  {\doibase 10.1103/PhysRevLett.119.050601} {\bibfield  {journal} {\bibinfo
  {journal} {Phys. Rev. Lett.}\ }\textbf {\bibinfo {volume} {119}},\ \bibinfo
  {pages} {050601} (\bibinfo {year} {2017})}\BibitemShut {NoStop}%
\bibitem [{\citenamefont {Scandi}\ and\ \citenamefont
  {Perarnau-Llobet}(2019)}]{Scandi}%
  \BibitemOpen
  \bibfield  {author} {\bibinfo {author} {\bibfnamefont {M.}~\bibnamefont
  {Scandi}}\ and\ \bibinfo {author} {\bibfnamefont {M.}~\bibnamefont
  {Perarnau-Llobet}},\ }\href@noop {} {\bibfield  {journal} {\bibinfo
  {journal} {Quantum}\ }\textbf {\bibinfo {volume} {3}},\ \bibinfo {pages}
  {197} (\bibinfo {year} {2019})}\BibitemShut {NoStop}%
\bibitem [{\citenamefont {Feldmann}\ and\ \citenamefont
  {Kosloff}(2003)}]{Feldmann}%
  \BibitemOpen
  \bibfield  {author} {\bibinfo {author} {\bibfnamefont {T.}~\bibnamefont
  {Feldmann}}\ and\ \bibinfo {author} {\bibfnamefont {R.}~\bibnamefont
  {Kosloff}},\ }\href@noop {} {\bibfield  {journal} {\bibinfo  {journal} {Phys.
  Rev. E}\ }\textbf {\bibinfo {volume} {68}},\ \bibinfo {pages} {016101}
  (\bibinfo {year} {2003})}\BibitemShut {NoStop}%
\bibitem [{\citenamefont {Braunstein}\ and\ \citenamefont {{Van
  Loock}}(2005)}]{Braunstein2005}%
  \BibitemOpen
  \bibfield  {author} {\bibinfo {author} {\bibfnamefont {L.~S.}\ \bibnamefont
  {Braunstein}}\ and\ \bibinfo {author} {\bibfnamefont {P.}~\bibnamefont {{Van
  Loock}}},\ }\href {\doibase 10.1103/RevModPhys.77.513} {\bibfield  {journal}
  {\bibinfo  {journal} {Rev. Mod. Phys.}\ }\textbf {\bibinfo {volume} {77}},\
  \bibinfo {pages} {513} (\bibinfo {year} {2005})}\BibitemShut {NoStop}%
\bibitem [{\citenamefont {Chen}(2005)}]{Chen2005}%
  \BibitemOpen
  \bibfield  {author} {\bibinfo {author} {\bibfnamefont {X.-y.}\ \bibnamefont
  {Chen}},\ }\href@noop {} {\bibfield  {journal} {\bibinfo  {journal} {Phys.
  Rev. A}\ }\textbf {\bibinfo {volume} {71}},\ \bibinfo {pages} {062320}
  (\bibinfo {year} {2005})}\BibitemShut {NoStop}%
\bibitem [{\citenamefont {Banchi}\ \emph {et~al.}(2015)\citenamefont {Banchi},
  \citenamefont {Braunstein},\ and\ \citenamefont {Pirandola}}]{Banchi2015}%
  \BibitemOpen
  \bibfield  {author} {\bibinfo {author} {\bibfnamefont {L.}~\bibnamefont
  {Banchi}}, \bibinfo {author} {\bibfnamefont {S.~L.}\ \bibnamefont
  {Braunstein}}, \ and\ \bibinfo {author} {\bibfnamefont {S.}~\bibnamefont
  {Pirandola}},\ }\href {\doibase 10.1103/PhysRevLett.115.260501} {\bibfield
  {journal} {\bibinfo  {journal} {Phys. Rev. Lett.}\ }\textbf {\bibinfo
  {volume} {115}},\ \bibinfo {pages} {260501} (\bibinfo {year}
  {2015})}\BibitemShut {NoStop}%
\bibitem [{\citenamefont {Tanaka}(2006)}]{Tanaka2006}%
  \BibitemOpen
  \bibfield  {author} {\bibinfo {author} {\bibfnamefont {F.}~\bibnamefont
  {Tanaka}},\ }\href {\doibase 10.1088/0305-4470/39/45/024} {\bibfield
  {journal} {\bibinfo  {journal} {J. Phys. A}\ }\textbf {\bibinfo {volume}
  {39}},\ \bibinfo {pages} {14165} (\bibinfo {year} {2006})}\BibitemShut
  {NoStop}%
\bibitem [{\citenamefont {Lenglet}\ \emph {et~al.}(2006)\citenamefont
  {Lenglet}, \citenamefont {Rousson}, \citenamefont {Deriche},\ and\
  \citenamefont {Faugeras}}]{Lenglet2006}%
  \BibitemOpen
  \bibfield  {author} {\bibinfo {author} {\bibfnamefont {C.}~\bibnamefont
  {Lenglet}}, \bibinfo {author} {\bibfnamefont {M.}~\bibnamefont {Rousson}},
  \bibinfo {author} {\bibfnamefont {R.}~\bibnamefont {Deriche}}, \ and\
  \bibinfo {author} {\bibfnamefont {O.}~\bibnamefont {Faugeras}},\ }in\ \href
  {\doibase 10.1007/s10851-006-6897-z} {\emph {\bibinfo {booktitle} {J. Math.
  Imag. Vis}}},\ Vol.~\bibinfo {volume} {25}\ (\bibinfo {year} {2006})\ pp.\
  \bibinfo {pages} {423--444}\BibitemShut {NoStop}%
\bibitem [{\citenamefont {Andai}(2009)}]{Andai2009a}%
  \BibitemOpen
  \bibfield  {author} {\bibinfo {author} {\bibfnamefont {A.}~\bibnamefont
  {Andai}},\ }\href {\doibase 10.1016/j.jmva.2008.08.007} {\bibfield  {journal}
  {\bibinfo  {journal} {J. Multi. Anal.}\ }\textbf {\bibinfo {volume} {100}},\
  \bibinfo {pages} {777} (\bibinfo {year} {2009})}\BibitemShut {NoStop}%
\bibitem [{\citenamefont {Feldmann}\ \emph {et~al.}(1985)\citenamefont
  {Feldmann}, \citenamefont {Andresen}, \citenamefont {Qi},\ and\ \citenamefont
  {Salamon}}]{Feldmann2013}%
  \BibitemOpen
  \bibfield  {author} {\bibinfo {author} {\bibfnamefont {T.}~\bibnamefont
  {Feldmann}}, \bibinfo {author} {\bibfnamefont {B.}~\bibnamefont {Andresen}},
  \bibinfo {author} {\bibfnamefont {A.}~\bibnamefont {Qi}}, \ and\ \bibinfo
  {author} {\bibfnamefont {P.}~\bibnamefont {Salamon}},\ }\href {\doibase
  10.1063/1.449666} {\bibfield  {journal} {\bibinfo  {journal} {J. Chem.
  Phys.}\ }\textbf {\bibinfo {volume} {83}},\ \bibinfo {pages} {5849} (\bibinfo
  {year} {1985})}\BibitemShut {NoStop}%
\bibitem [{\citenamefont {Pinele}\ \emph {et~al.}(2020)\citenamefont {Pinele},
  \citenamefont {Strapasson},\ and\ \citenamefont {Costa}}]{Distributions2020}%
  \BibitemOpen
  \bibfield  {author} {\bibinfo {author} {\bibfnamefont {J.}~\bibnamefont
  {Pinele}}, \bibinfo {author} {\bibfnamefont {J.~E.}\ \bibnamefont
  {Strapasson}}, \ and\ \bibinfo {author} {\bibfnamefont {S.~I.~R.}\
  \bibnamefont {Costa}},\ }\href {\doibase 10.3390/e22040404} {\bibfield
  {journal} {\bibinfo  {journal} {Entropy}\ }\textbf {\bibinfo {volume} {22}},\
  \bibinfo {pages} {404} (\bibinfo {year} {2020})}\BibitemShut {NoStop}%
\bibitem [{\citenamefont {Solon}\ and\ \citenamefont
  {Horowitz}(2018)}]{Solon2018a}%
  \BibitemOpen
  \bibfield  {author} {\bibinfo {author} {\bibfnamefont {A.~P.}\ \bibnamefont
  {Solon}}\ and\ \bibinfo {author} {\bibfnamefont {J.~M.}\ \bibnamefont
  {Horowitz}},\ }\href@noop {} {\bibfield  {journal} {\bibinfo  {journal}
  {Phys. Rev. Lett.}\ }\textbf {\bibinfo {volume} {120}},\ \bibinfo {pages}
  {180605} (\bibinfo {year} {2018})}\BibitemShut {NoStop}%
\bibitem [{\citenamefont {Schmiedl}\ and\ \citenamefont
  {Seifert}()}]{Schmiedl}%
  \BibitemOpen
  \bibfield  {author} {\bibinfo {author} {\bibfnamefont {T.}~\bibnamefont
  {Schmiedl}}\ and\ \bibinfo {author} {\bibfnamefont {U.}~\bibnamefont
  {Seifert}},\ }\href@noop {} {\bibfield  {journal} {\bibinfo  {journal} {Phys.
  Rev. Lett.}\ }\textbf {\bibinfo {volume} {98}},\ \bibinfo {pages}
  {108301}}\BibitemShut {NoStop}%
\bibitem [{\citenamefont {Blaber}\ \emph {et~al.}(2021)\citenamefont {Blaber},
  \citenamefont {Louwerse},\ and\ \citenamefont {Sivak}}]{Blaber2021}%
  \BibitemOpen
  \bibfield  {author} {\bibinfo {author} {\bibfnamefont {S.}~\bibnamefont
  {Blaber}}, \bibinfo {author} {\bibfnamefont {M.~D.}\ \bibnamefont
  {Louwerse}}, \ and\ \bibinfo {author} {\bibfnamefont {D.~A.}\ \bibnamefont
  {Sivak}},\ }\href {http://arxiv.org/abs/2105.04691} {\bibfield  {journal}
  {\bibinfo  {journal} {Phys. Rev. E}\ }\textbf {\bibinfo {volume} {104}},\
  \bibinfo {pages} {022101} (\bibinfo {year} {2021})}\BibitemShut {NoStop}%
\end{thebibliography}%

\end{document}